\documentclass[draft,11pt]{article}
\usepackage{amssymb,amsmath,theorem,a4wide}

\allowdisplaybreaks[2]

\newcommand\M{\mathit}

\newcommand\eq{\leftrightarrow}
\renewcommand\iff{\quad\text{iff}\quad}
\newcommand\LOR{\bigvee}
\newcommand\ET{\bigwedge}
\newcommand\TO{\Rightarrow}
\newcommand\nneg{\mathord\sim}
\newcommand\ieq{\rightleftarrows}
\newcommand\model{\vDash}
\newcommand\nmodel{\nvDash}

\newcommand\fii{\varphi}
\newcommand\roo{\varrho}

\newcommand\p[1]{\langle#1\rangle}
\newcommand\lh[1]{\lvert#1\rvert}
\newcommand\Abs[1]{\left|#1\right|}
\newcommand\bez{\smallsetminus}
\newcommand\sset{\subseteq}
\newcommand\nsset{\nsubseteq}
\newcommand\Sset{\supseteq}
\newcommand\pw[1]{\mathcal P(#1)}
\newcommand\nul{\varnothing}
\newcommand\res{\mathbin\restriction}

\newcommand\ru{\mathrel/}
\newcommand\up{\mathord\uparrow}

\DeclareMathOperator\id{id}
\DeclareMathOperator\Form{Form}
\DeclareMathOperator\limd{ld_\to}

\newcommand\lgc{\mathbf}
\newcommand\EF[1]{#1\text-\M{EF}}
\newcommand\SF[1]{#1\text-\M{SF}}
\newcommand\CF[1]{#1\text-\M{CF}}
\newcommand\Fr[1]{#1\text-\M{F}}
\newcommand\CPC{\lgc{CPC}}
\newcommand\IPC{\lgc{IPC}}
\newcommand\bckw{\mathrm{BCKW}}
\newcommand\I{{\bullet}}
\newcommand\R{{\circ}}
\newcommand\clsop{\operatorname}

\newcommand\cF{\mathcal F}
\newcommand\cM{\mathcal M}

\newcommand\ob{\overline}

\def\cput(#1)#2{\put(#1){\hbox to0pt{\hss#2\hss}}}

\newcommand\bme{\hskip.75em\relax}
\newcommand\noproof{\leavevmode\unskip\bme\vadjust{}\nobreak\hfill$\qed$\par}
\newcommand\qed{\Box}
\newenvironment{Pf}
  {\par\noindent\textit{Proof:}\bme\ignorespaces}
  {\noproof\pagebreak[2]\vskip\medskipamount\ignorespacesafterend}
\newcommand\qedhere{\relax\ifmmode\eqno\qed\expandafter\aftergroup
                   \else\noproof\fi\noqed}
\newcommand\noqed{\let\noproof\relax}

\theoremstyle{plain}
\newtheorem{Thm}{Theorem}[section]
\newtheorem{Prop}[Thm]{Proposition}
\newtheorem{Cor}[Thm]{Corollary}
\newtheorem{Lem}[Thm]{Lemma}
\newtheorem{Obs}[Thm]{Observation}
\newtheorem{Que}[Thm]{Question}
\newtheorem{FThm}[Thm]{False theorem}
\newtheorem{Cl}{Claim}[Thm]
\renewcommand\theCl{\arabic{Cl}}

\theorembodyfont\upshape
\newtheorem{Def}[Thm]{Definition}
\newtheorem{Rem}[Thm]{Remark}
\newtheorem{Exm}[Thm]{Example}
\newenvironment{Pf*}{\let\qed\qedCl\Pf}\endPf

\author{Emil Je\v r\'abek\\[\medskipamount]
Institute of Mathematics of the Czech Academy of Sciences\\
\small \v Zitn\'a 25,
115\:67 Praha 1,
Czech Republic,
email: \texttt{jerabek@math.cas.cz}
}

\title{Proof complexity of intuitionistic implicational formulas}

\begin{document}
\maketitle

\begin{abstract}
We study implicational formulas in the context of proof complexity of intuitionistic propositional logic ($\IPC$). On the one hand, we give an
efficient transformation of tautologies to implicational tautologies that preserves the lengths of intuitionistic
extended Frege ($\M{EF}$) or substitution Frege ($\M{SF}$) proofs up to a polynomial. On the other hand,
$\M{EF}$~proofs in the implicational fragment of~$\IPC$
polynomially simulate full intuitionistic logic for implicational tautologies. The results also apply to other fragments of other superintuitionistic
logics under certain conditions.

In particular, the exponential lower bounds on the length of intuitionistic $\M{EF}$~proofs by Hrube\v
s~\cite{hru:lbint}, generalized to exponential separation between $\M{EF}$ and~$\M{SF}$ systems in superintuitionistic
logics of unbounded branching by Je\v r\'abek~\cite{ej:sfef}, can be realized by implicational tautologies.


\end{abstract}

\section{Introduction}

A major open problem in proof complexity is to show superpolynomial lower bounds on the lengths of proofs in Frege
systems for classical propositional logic, or even stronger systems such as extended Frege. It turns out such lower
bounds are easier to obtain for some non-classical logics: Hrube\v s proved exponential lower bounds on the length of
$\M{EF}$~proofs%
\footnote{The results are formulated in~\cite{hru:lbmod,hru:lbint,hru:nonclas} as lower bounds on the number of lines
in Frege proofs, however, this is essentially the same measure as the length of extended Frege proofs.}
for certain modal logics and for intuitionistic logic~\cite{hru:lbmod,hru:lbint,hru:nonclas}.
Je\v r\'abek~\cite{ej:sfef} improved these results to an exponential separation between $\M{EF}$ and~$\M{SF}$ systems
for all superintuitionistic (and transitive modal) logics of unbounded branching. See also \cite{buss-mints,buss-pud}
for earlier work on the proof complexity of intuitionistic logic, including conditional lower bounds.

Known lower bounds on proof systems for non-classical logics crucially rely on variants of the feasible disjunction
property, serving a similar role as feasible interpolation does in weak proof systems for classical logic (cf.\
\cite{kra:fi}).
Consequently, the lower bounds are proved for tautologies that involve disjunction in an essential way, and one might
get the impression that this is unavoidable---perhaps the implicational fragment of~$\IPC$, or other disjunction-free
fragments, behave differently from the full logic.

The purpose of this paper is to show that this dependence on disjunctions is just an artifact of the proofs: the
implicational fragment of intuitionistic logic is, after all, essentially equivalent to the full logic with respect to
the lengths of proofs. We will demonstrate this by means of two kinds of results: first, \emph{tautologies} can be
brought into a form avoiding unwanted connectives (such as disjunction) while preserving their hardness for
intuitionistic extended Frege and related systems; second, unwanted connectives can be eliminated from intuitionistic
extended Frege \emph{proofs} except for subformulas of the tautology being proved. We include several results of both
kinds with varying assumptions.

Elimination results of the first kind are the topic of Section~\ref{sec:elim-taut}. On the one hand, in
Theorem~\ref{prop:impessall} we present a method that makes tautologies \emph{mostly} implicational with certain
disjunctions and~$\bot$ left, and preserves (up to a polynomial) the size of $F$, $\M{EF}$, and $\M{SF}$ proofs in
\emph{arbitrary} superintuitionistic logics; in particular, the tautologies with exponential $\M{EF}$ lower bounds from
\cite{hru:lbint,ej:sfef} can be made purely implicational in this way. On the other hand, in
Theorems \ref{prop:bot-free} and~\ref{prop:lor-free} we show how to eliminate \emph{all} disjunctions (and/or $\bot$) from tautologies
while preserving the lengths of $\M{EF}$ and $\M{SF}$ proofs in logics whose proper axioms do not contain disjunctions
($\bot$, respectively).

Elimination results of the second kind come in Section~\ref{sec:elim-proof}. In Theorem~\ref{prop:lor-bot-pfs}, we show that
if the proper axioms of a logic~$L$ do not contain disjunction (or $\bot$), we can efficiently eliminate disjunctions
($\bot$, resp.) from $\EF L$ proofs, except for those that appear in the final tautology. However, the argument may
introduce conjunctions, and we address this in subsequent results: in Corollary~\ref{cor:all-elim} and Theorem~\ref{thm:all-but-bot}, we show how
to eliminate conjunctions from $\M{EF}$-proofs under some conditions on the logic and its axiom system; in
Theorem~\ref{thm:bot-top}, we show how to eliminate $\bot$ from proofs without introducing $\land$ or other connectives,
again under certain conditions on the logic. We also develop a monotone version of the negative translation
(Proposition~\ref{lem:neg}), which we use in an ad hoc argument that the  above-mentioned implicational versions of the
tautologies used in \cite{ej:sfef} to separate $\M{EF}$ and~$\M{SF}$ have short implicational $\SF\IPC$ proofs
(Theorem~\ref{thm:sep-imp}).

A few concluding remarks and open problems are mentioned in Section~\ref{sec:conclusion}.

In order to show the limitations of our methods, the appendices include some negative results that may be of
independent interest. Proposition~\ref{prop:frag}, originally due to Wro\'nski~\cite{wron:red}, shows that in general, the
implicational fragment of a superintuitionistic logic $L=\IPC+\Phi$ with $\Phi$ an implicational axiom may not be
axiomatized by~$\Phi$ over the implicational fragment of~$\IPC$, and similarly for other combinations where the target
fragment omits conjunction. Appendix~\ref{sec:cxt-flas} presents certain exponential lower bounds on the size of
formulas in fragments of intuitionistic logic, and a linear lower bound on implication nesting depth.

\section{Preliminaries}

We refer to Chagrov and Zakharyaschev~\cite{cha-zax} and Kraj\'\i\v cek~\cite{book} for general information on
superintuitionistic logics and classical Frege systems and their extensions, respectively. For Frege and friends
in superintuitionistic logics, we will use the notation and basic results from Je\v r\'abek~\cite{ej:sfef}; we include
more details below, as we need to generalize the set-up to fragments (e.g., implicational) of si logics, which were not
treated in~\cite{ej:sfef}.

If $L$ is a language\footnote{In the sense of the theory of computation, i.e., an arbitrary set of strings over a
finite alphabet.}, a \emph{proof system for~$L$} is a polynomial-time computable function $P(x)$
whose range is~$L$. If $P(x)=y$, the string~$x$ is called a \emph{$P$-proof} of~$y$. A proof system $P$
\emph{polynomially-simulates} or \emph{p-simulates} a proof system~$Q$, denoted $Q\le_pP$, if there is a poly-time function $f$ such that
$Q(x)=P(f(x))$ for all proofs~$x$. Proof systems $P$ and~$Q$ are \emph{p-equivalent}, written $P\equiv_pQ$, if
$P\le_pQ\le_pP$.

A \emph{propositional language} is a set~$C$ of connectives, each given a finite arity. (That is, formally, a language
is a mapping $\mathit{ar}\colon C\to\omega$.) Let $\Form_C$ denote the set of
\emph{$C$-formulas}, built from propositional variables $p_i$, $i\in\omega$, using connectives from~$C$. We will also
denote variables by other lowercase Latin letters for convenience, and we will denote formulas by lowercase Greek
letters. We write $\psi\sset\fii$ if $\psi$ is a subformula of~$\fii$. A \emph{substitution} is a mapping of variables to formulas, uniquely extended to a homomorphism
$\Form_C\to\Form_C$. A \emph{propositional logic} is a pair $L=\p{C,{\vdash}_L}$, where $C$ is a propositional
language, and $\vdash_L$ is a structural Tarski-style consequence relation: i.e., a relation
${\vdash}_L\sset\pw{\Form_C}\times\Form_C$ satisfying
\begin{enumerate}
\item $\fii\in\Gamma$ implies $\Gamma\vdash_L\fii$;
\item if $\Delta\vdash_L\fii$, and $\Gamma\vdash_L\psi$ for all $\psi\in\Delta$, then $\Gamma\vdash_L\fii$;
\item $\Gamma\vdash_L\fii$ implies $\sigma(\Gamma)\vdash_L\sigma(\fii)$
\end{enumerate}
for all formulas $\fii$, sets of formulas $\Gamma$, $\Delta$, and substitutions~$\sigma$. We will only be interested in
\emph{finitary} logics, meaning $\Gamma\vdash_L\fii$ implies $\Gamma_0\vdash_L\fii$ for some finite
$\Gamma_0\sset\Gamma$.

If $L$ is a logic in a finite (or computable) language~$C$, a \emph{proof system for~$L$} is a proof system (in the
sense given above) for the set $\{\fii:{}\vdash_L\fii\}$ of \emph{$L$-tautologies}; a \emph{derivation system for~$L$}
is a proof system for its finitary consequence relation
$\{\p{\fii_0,\fii_1,\dots,\fii_n}:\fii_1,\dots,\fii_n\vdash_L\fii_0\}$.

Let $L=\p{C,{\vdash}_L}$ be a logic. An \emph{extension} of $L$ is a logic $L'=\p{C',{\vdash}_{L'}}$ such that $C\sset
C'$ and ${\vdash}_L\sset{\vdash}_{L'}$; we will write $L\sset L'$. An extension is \emph{simple} if $C=C'$. If $X$ is a
set of $C$-formulas, then $L+X$ is the least extension of~$L$ such that $\vdash_{L+X}\fii$ for all $\fii\in X$.
Logics of the form $L+X$ are called \emph{axiomatic extensions} of~$L$. For any $C'\sset C$, the \emph{$C'$-fragment
of~$L$}, denoted~$L_{C'}$, is the logic $\p{C',{\vdash}_L\cap(\pw{\Form_{C'}}\times\Form_{C'})}$.

Fix a language~$C$. A \emph{Frege rule} is an object of the form
\[\frac{\alpha_1,\dots,\alpha_k}{\alpha_0},\qquad\text{also written as}\quad
\alpha_1,\dots,\alpha_k\ru\alpha_0,\]
where $\alpha_0,\dots,\alpha_k$ are formulas. Rules with $k=0$ can be identified with formulas, and are called
\emph{axioms}. If $L$ is a logic, a rule $\alpha_1,\dots,\alpha_k\ru\alpha_0$ is \emph{$L$-derivable} if $\alpha_1,\dots,\alpha_k\vdash_L\alpha_0$.

If $R$ is a set of Frege rules, an \emph{$R$-derivation} of
$\fii$ from $\Gamma\sset\Form_C$ is a sequence of formulas $\fii_1,\dots,\fii_m$, where $\fii_m=\fii$, and each
$\fii_i$ is an element of~$\Gamma$, or is derived from some of the
formulas $\fii_1,\dots,\fii_{i-1}$ by a substitution instance of a rule from~$R$. An \emph{$R$-proof} of $\fii$ is its $R$-derivation from~$\Gamma=\nul$. If $L$ is
a logic such that 
\[\Gamma\vdash_L\fii\iff\text{$\fii$ has an $R$-derivation from $\Gamma$}\]
for all $\Gamma\cup\{\fii\}\sset\Form_C$, we say that $R$ is an \emph{axiomatization} of~$L$; if $R$ is furthermore
finite, we call the $R$-derivation system a \emph{Frege system for~$L$}, and denote it $\Fr L$. This notation is
justified by the next observation:
\begin{Lem}\label{lem:equi}
Substitution instances of a fixed $L$-derivable rule have linear-time constructible proofs in any $\Fr L$ system.
Consequently, all Frege systems for~$L$ are p-equivalent.
\noproof\end{Lem}
Notice that our definition is more strict than the one employed in~\cite{min-koj,ej:frege}, which only requires Frege
systems to respect the tautologies of the logic rather than its consequence relation.

The formulas $\fii_i$ in a Frege derivation $\fii_1,\dots,\fii_m$ are called \emph{lines}; thus, the number of lines in the
derivation is~$m$, whereas its \emph{size} or \emph{length} is its total size when written out as a string in a fixed
finite alphabet, i.e., essentially $\sum_i\lh{\fii_i}$.

The logics we will work with in this paper are extensions of intuitionistic logic (denoted $\IPC$) and their fragments.
We formulate $\IPC$ in the language $C_\IPC=\{\to,\land,\lor,\bot\}$. We define $\neg\fii=(\fii\to\bot)$ as an abbreviation.
Conversely, $\bot$ is equivalent to $\neg(p\to p)$; since implication will always be present in our fragments, we could
equally well formulate all results using $\neg$ instead of~$\bot$ as a basic connective, but we feel $\bot$ is more
convenient. We can also define $\top$ as $\bot\to\bot$ or $p\to p$, and
$(\fii\eq\psi)=(\fii\to\psi)\land(\psi\to\fii)$. We will call $\{\to\}$-formulas \emph{implicational} (or \emph{purely
implicational} for emphasis), $\{\to,\land,\lor\}$-formulas
\emph{positive}, $\{\land,\lor,\bot,\top\}$-formulas \emph{monotone}, and $\{\land,\lor\}$-formulas \emph{strict
monotone}. (Every monotone formula is equivalent to $\bot$, $\top$, or to a strict monotone formula.)

In order not to get overwhelmed by brackets, we employ the notational conventions that the outermost bracket can be
omitted, $\land$ and~$\lor$ bind more strongly than $\to$, and $\to$ is right-associative, so that e.g.,
\[\fii_0\to\dots\to\fii_{n-1}\to\psi\land\chi=(\fii_0\to(\cdots\to(\fii_{n-1}\to(\psi\land\chi))\cdots)).\]
If $\Gamma=\p{\fii_i:i<n}$ is a sequence of formulas, we will write
\[\Gamma\to\fii=\fii_0\to\dots\to\fii_{n-1}\to\fii,\]
understood as just~$\fii$ if $n=0$. Any formula~$\fii$ can be uniquely written in the form $\Gamma\to\xi$, where
$\Gamma$ is a sequence of formulas, and $\xi$ is not an implication. We will call $\xi$ the \emph{head} of~$\fii$, and
write $H(\fii)=\xi$.

We can axiomatize~$\IPC$ by the rule of modus ponens
\begin{equation}\label{eq:41}
\alpha,\alpha\to\beta\ru\beta,
\end{equation}
and the axioms
\begin{gather}
\label{eq:18}(\alpha\to\beta\to\gamma)\to(\alpha\to\beta)\to\alpha\to\gamma,\\
\label{eq:31}\alpha\to\beta\to\alpha,\\
\label{eq:32}\alpha\land\beta\to\alpha,\\
\label{eq:33}\alpha\land\beta\to\beta,\\
\label{eq:30}\alpha\to\beta\to\alpha\land\beta,\\
\label{eq:11}\alpha\to\alpha\lor\beta,\\
\label{eq:12}\beta\to\alpha\lor\beta,\\
\label{eq:13}\alpha\lor\beta\to(\alpha\to\gamma)\to(\beta\to\gamma)\to\gamma,\\
\label{eq:7}\bot\to\gamma.
\end{gather}
The Frege system for~$\IPC$ based on the rules \eqref{eq:41}--\eqref{eq:7} will be called the \emph{standard} $\Fr\IPC$
system. 

\emph{Superintuitionistic (si)} logics are axiomatic extensions of~$\IPC$. The largest consistent si~logic is classical
logic, which we will denote~$\CPC$. Only finitely axiomatizable logics have Frege systems, hence unless stated
otherwise, we will tacitly assume that all logics mentioned are finitely axiomatizable.

We will rarely need to use Kripke semantics for si logics, as most of our arguments are purely syntactic; we refer the
reader to \cite{cha-zax,ej:sfef}.

Apart from si logics themselves, we will be interested in proof systems for fragments $L_C$, where $L$ is a si logic,
and ${\to}\in C\sset C_\IPC$. (We warn the reader that fragments of finitely axiomatized logics are not necessarily
finitely axiomatizable, and axiomatic extensions of $\IPC_C$ are not necessarily fragments of si logics.) Notice that
if finitely axiomatizable, $L_C$ can be axiomatized by a single formula over~$\IPC_C$: in particular,
$\IPC_C+\{\fii_1,\dots,\fii_k\}$ can be axiomatized by
\[(\fii_1\to\dots\to\fii_k\to p)\to p,\]
where $p$ is a variable not occurring in $\fii_1,\dots,\fii_k$. It is well known (see e.g.\ \cite{horn:sep-int}, using
a slightly different calculus) that the standard axiom system
for~$\IPC$ above has the property of \emph{separation of connectives} for fragments that include implication: $\IPC_C$ is
axiomatizable by modus ponens and those axioms among \eqref{eq:18}--\eqref{eq:7} that only use connectives from~$C$.
Thus, an $\Fr{L_C}$ system axiomatized by \eqref{eq:41}, the $C$-axioms among \eqref{eq:18}--\eqref{eq:7}, and one
other $C$-formula (called the \emph{proper axiom} of the given axiom system) will be called a \emph{standard Frege system for~$L_C$}.

We mention that si logics split in four types according to which connectives need to appear in their proper axioms.
First, no consistent proper extension of~$\IPC$ can be axiomatized by monotone formulas, hence $\to$ is (nearly) always
required. On the other hand, conjunction is never needed:
\begin{Lem}\label{lem:conj-free}
Let $C\sset\{\to,\lor,\bot\}$. Every $(C\cup\{\land\})$-formula is in $\IPC$ equivalent to a conjunction of $C$-formulas.
\end{Lem}
\begin{Pf}
By induction on the complexity of the formula, using
\begin{align}
\Bigl(\ET_{i<n}\fii_i\Bigr)\lor\Bigl(\ET_{j<m}\psi_j\Bigr)&\eq
  \ET_{\substack{i<n\\j<m}}(\fii_i\lor\psi_j),\\
\Bigl(\ET_{i<n}\fii_i\to\ET_{j<m}\psi_j\Bigr)&\eq
  \ET_{j<m}(\fii_0\to\dots\to\fii_{n-1}\to\psi_j).
\end{align}
\end{Pf}
Thus, the only question is whether axioms of~$L$ need to use $\lor$ and/or $\bot$, and this can be characterized
semantically (see \cite{cha-zax} for a detailed explanation): $L$ can be axiomatized by $\{{\to},\lor\}$-formulas (or:
by positive formulas) iff the class of $L$-frames is closed under so-called dense subframes (cf.~\cite[\S9.1]{cha-zax}); $L$ is axiomatizable
by $\{\to,\bot\}$-formulas (or: by $\lor$-free formulas) iff it is a \emph{cofinal-subframe logic} (cf.\ \cite[\S11.3]{cha-zax}); and $L$ is axiomatizable by
$\to$-formulas (or: by positive $\lor$-free formulas) iff it is a \emph{subframe logic} iff it is both positively
axiomatizable and cofinal-subframe.

For any Frege system $\Fr L$, we will also consider several associated stronger systems based on the same set of Frege
rules. The \emph{extended Frege} system $\EF L$ is defined as follows: an $\EF L$ derivation of $\fii$ from a finite
set~$\Gamma$ is an $\Fr L$ derivation of $\fii$ from $\Gamma\cup E$, where $E=\{E_1,\dots,E_r\}$ is a list of
\emph{extension axioms}: each $E_i$ is a pair of implications%
\footnote{This formulation of $\M{EF}$ is tailored to fragments of si logics. In general, we can reasonably define
$\M{EF}$ for (finitary, finitely axiomatized) logics that are \emph{finitely equivalential} in the sense of abstract
algebraic logic, where the set of equivalence formulas will play the role of $\{x\to y,y\to x\}$ in extension axioms.}
\[q_i\to\psi_i,\qquad\psi_i\to q_i,\]
where the \emph{extension variables} $q_i$ are distinct, and $q_i$ does not occur in $\{\psi_j:j\le
i\}\cup\{\fii\}\cup\Gamma$.
Alternatively, the \emph{circuit Frege} system $\CF L$ is defined similarly to Frege, except that lines in derivations can be
represented by circuits rather than just formulas.
In the \emph{substitution Frege} system $\SF L$, substitution can be used as a rule of inference along with Frege
rules. Unlike the other systems, $\SF L$ only makes sense as a proof system, not as a derivation system. If $P$ is any
of the sequence-like (dag-like) proof systems we introduced, $P^*$ denotes its tree-like version.

Lemma~\ref{lem:equi} also applies to $\EF L$, $\CF L$, and $\SF L$, but the case of tree-like systems is more complicated
(see \cite[\S3.1]{ej:sfef}).

Before embarking on the basic properties of these systems, let us mention a technical lemma that allows us to work with
chained implications $\Gamma\to\fii$ with more ease. It is reasonably clear that the formulas below have short
$\Fr\IPC$ proofs, but it may be less obvious how to derive them shortly in the implicational fragment $\IPC_\to$ without the
use of conjunctions.
\pagebreak[2]
\begin{Lem}\label{lem:struct}
\ \begin{enumerate}
\item\label{item:12}
Given $\Gamma$, $\Delta$, and $\fii$ such that every formula in~$\Gamma$ appears in~$\Delta$, we can construct in
polynomial time an $\Fr{\IPC_\to}$ proof of
\[(\Gamma\to\fii)\to(\Delta\to\fii).\]
\item\label{item:13}
Given $\Gamma$, $\fii$, and $\Delta=\p{\psi_i:i<n}$, we can construct in polynomial time an $\Fr{\IPC_\to}$ proof of
\[(\Delta\to\fii)\to(\Gamma\to\psi_0)\to\dots\to(\Gamma\to\psi_{n-1})\to(\Gamma\to\fii).\]
\end{enumerate}
\end{Lem}
\begin{Pf}
Instances of
\begin{equation}\label{eq:44}
(\psi\to\fii)\to(\gamma\to\psi)\to(\gamma\to\fii)
\end{equation}
have short $\Fr{\IPC_\to}$ proofs by Lemma~\ref{lem:equi}. By chaining $\lh\Gamma$ of them, we can construct short proofs
of
\begin{equation}\label{eq:45}
(\psi\to\fii)\to(\Gamma\to\psi)\to(\Gamma\to\fii).
\end{equation}
(This is in fact the special case of~\ref{item:13} with $n=1$.)

\ref{item:12}: It is easy to see that we can derive each such formula as a composition of polynomially many instances
of weakening, exchange, and contraction:
\begin{gather}
\label{eq:21}(\Gamma\to\Pi\to\fii)\to(\Gamma\to\alpha\to\Pi\to\fii),\\
\label{eq:42}(\Gamma\to\alpha\to\beta\to\Pi\to\fii)\to(\Gamma\to\beta\to\alpha\to\Pi\to\fii),\\
\label{eq:43}(\Gamma\to\alpha\to\alpha\to\Pi\to\fii)\to(\Gamma\to\alpha\to\Pi\to\fii),
\end{gather}
it thus suffices to prove \eqref{eq:21}--\eqref{eq:43}. By replacing $\fii$ with $\Pi\to\fii$, we can assume
$\Pi=\nul$; using \eqref{eq:45}, we can also assume $\Gamma=\nul$. Then \eqref{eq:21}--\eqref{eq:43} are instances of
constant-size $\IPC_\to$ tautologies, hence they have short $\Fr{\IPC_\to}$ proofs by Lemma~\ref{lem:equi}.

\ref{item:13}: By induction on~$n$. For $n=0$, \ref{item:13} reduces to $\fii\to\Gamma\to\fii$, which is an instance
of~\ref{item:12}. For the induction step, let $\Delta=\p{\psi_0,\dots,\psi_n}$, and put
$\Delta'=\p{\psi_1,\dots,\psi_n}$, and
\[\xi=(\Gamma\to\psi_1)\to\dots\to(\Gamma\to\psi_n)\to\Gamma\to\fii.\]
By the induction hypothesis, we have an $\Fr{\IPC_\to}$ proof of
\[(\Delta'\to\fii)\to\xi.\]
Using~\eqref{eq:44}, we infer
\[(\psi_0\to\Delta'\to\fii)\to(\psi_0\to\xi),\]
and using~\eqref{eq:45}, we obtain
\[(\psi_0\to\Delta'\to\fii)\to(\Gamma\to\psi_0)\to\Gamma\to\xi,\]
that is,
\[(\Delta\to\fii)\to(\Gamma\to\psi_0)\to\Gamma\to(\Gamma\to\psi_1)\to\dots\to(\Gamma\to\psi_n)\to\Gamma\to\fii.\]
We conclude
\[(\Delta\to\fii)\to(\Gamma\to\psi_0)\to(\Gamma\to\psi_1)\to\dots\to(\Gamma\to\psi_n)\to\Gamma\to\fii\]
using~\ref{item:12}.
\end{Pf}
As a consequence of~\ref{item:12}, we can extend the notation $\Gamma\to\fii$ to \emph{sets~$\Gamma$} instead of sequences, as
the ordering of the formulas does not make a significant difference. A particular special case of \ref{item:13}
is when $\vdash\Delta\to\fii$; for example, we can construct in poly-time $\Fr{\IPC_\to}$ proofs of
\[(\Gamma\to\fii)\to(\Gamma\to\fii\to\psi)\to\Gamma\to\psi.\]

Until the end of this section, let $L$ be an si logic, and $C\sset C_\IPC$ a set of connectives such that ${\to}\in
C$. Armed with Lemma~\ref{lem:struct}, we can prove the \emph{feasible deduction theorem} (cf.~\cite[Prop.~3.6]{ej:sfef})
using the standard argument by induction on the length of the derivation:
\begin{Prop}\label{prop:deduction}
Let $P$ denote $\Fr{L_C}$, $\CF{L_C}$, or $\EF{L_C}$. Given a $P$-derivation of a formula~$\fii$ from a set~$\Gamma$,
we can construct in polynomial time a $P$-proof of\/ $\Gamma\to\fii$.
\noproof\end{Prop}

Now we turn to the relations between our proof systems. First, $\CF L$ and $\EF L$ are more-or-less just different
presentations of the same system; moreover, the \emph{size} of $\EF L$ or $\CF L$ proofs is (up to a polynomial)
essentially the same measure as the number of \emph{lines} in $\Fr L$ proofs. These properties are well known for
$L=\CPC$, and noted to hold for si logics in \cite[Prop.~3.2--3]{ej:sfef}; we observe that the argument applies to
fragments~$L_C$ just the same. In order to state it properly, let $s_P(\fii)$ denote the minimal size of a $P$-proof of
a formula~$\fii$, and $k_P(\fii)$ the minimal number of lines in a $P$-proof of~$\fii$ (for proof systems $P$ where the
concept of lines makes sense).
\begin{Prop}\label{prop:efcf}
$\EF{L_C}$ and $\CF{L_C}$ are p-equivalent. Moreover, for any $L_C$-tautology $\fii$, we have
\[k_{\EF{L_C}}(\fii)\le k_{\Fr{L_C}}(\fii)=k_{\CF{L_C}}(\fii)=O(k_{\EF{L_C}}(\fii)),\]
and
\[s_{\EF{L_C}}(\fii)=O(k_{\Fr{L_C}}(\fii)+\lh\fii^2).\qedhere\]
\end{Prop}

The next result is again well known for $\CPC$, and noted to hold for si logics in
\cite[Prop.~3.11]{ej:sfef}. Even with Lemma~\ref{lem:struct} and Proposition~\ref{prop:deduction} at hand, it is not immediately obvious that it also applies to fragments, since the standard argument from
\cite[L.~4.4.8]{book} relies on a tree of conjunctions. We can make it work with the help of the following trick: in
intuitionistic logic, propositions expressed by formulas of the form $\Gamma\to\alpha$ for a fixed~$\alpha$ form a
Boolean algebra, with the induced $\to$~operation, and $\alpha$ serving the role of~$\bot$; cf.\ \eqref{eq:19}. In
particular, they carry a well-behaved conjunction operation
\[(\Gamma\to\alpha)\land^\alpha(\Delta\to\alpha)=\bigl((\Gamma\to\alpha)\to(\Delta\to\alpha)\to\alpha\bigr)\to\alpha.\]
This can be employed to emulate the original Kraj\'\i\v cek's argument. We leave the details to the reader.
\begin{Prop}\label{prop:tree-like}
Let $P$ be $\Fr{L_C}$, $\EF{L_C}$, or $\CF{L_C}$. Given a $P$-proof of a formula $\fii$ with $m$~lines,
we can construct in polynomial time a $P$-proof of $\fii$ of height $O(\log m)$. Consequently, $P\equiv_pP^*$.
\noproof\end{Prop}
Another consequence of Proposition~\ref{prop:tree-like} is that all tree-like Frege ($\M{EF}$, $\M{CF}$, resp.) systems
for~$L_C$ are equivalent. This is no longer true for tree-like $\M{SF}$ systems: as explained in \cite[\S3.1]{ej:sfef},
$\SF{L_C}^*$ systems are sensitive to the choice of basic rules\footnote{Thus, the $\SF{L_C}^*$ notation is
misleading. We should properly indicate the set of Frege rules.}. For simplicity, we state the next result for
\emph{standard} axiom systems, but it 
holds more generally under the assumption that \eqref{eq:41} has a tree-like Frege derivation in which the premise
$\alpha$ is used only once. Then, using Lemma~\ref{lem:struct} and Propositions \ref{prop:deduction} and~\ref{prop:tree-like}, we can generalize
\cite[Thm.~3.12]{ej:sfef} to fragments as follows:
\begin{Prop}
A standard $\SF{L_C}^*$ system is p-equivalent to $\EF{L_C}$.
\noproof\end{Prop}

This leaves us with only three potentially different proof systems for~$L_C$ out of those mentioned above (ignoring
non-standard $\SF{L_C}^*$ systems): $\Fr{L_C}\le_p\EF{L_C}\le_p\SF{L_C}$. Whether
$\EF\CPC$ has superpolynomial speed-up over $\Fr\CPC$ is a notorious open problem, and it is unresolved for $\IPC$ or
other si logics or their fragments as well. Whether $\M{EF}$ and $\M{SF}$ are equivalent depends on the logic: in particular,
$\EF\CPC\equiv_p\SF\CPC$, whereas $\SF\IPC$ has exponential speed-up over $\EF\IPC$. These questions are extensively
studied in \cite{ej:sfef}, while generalization of the speed-up to fragments is one of the goals of the present paper, so
we will get back to it later.

\section{Elimination of connectives from tautologies}\label{sec:elim-taut}

As already mentioned in the introduction, motivation for this section comes from lower bounds on intuitionistic
proofs:
\begin{Thm}[Hrube\v s \cite{hru:lbint}]\label{thm:hru-lb}
There is a sequence $\{\fii_n:n\in\omega\}$ of intuitionistic tautologies of the form
\begin{equation}\label{eq:1}
\ET_{i<m}(p_i\lor p'_i)\to\neg\alpha(\vec p,\vec s)\lor\neg\beta(\vec{p'},\vec r),
\end{equation}
constructible in time $n^{O(1)}$, that require $\EF\IPC$ proofs of size $2^{n^{\Omega(1)}}$.
\noproof\end{Thm}
The original clique--colouring tautologies can be modified to yield the following
strengthening:
\begin{Thm}[Je\v r\'abek \cite{ej:sfef}]\label{thm:sfef-lb}
There is a sequence $\{\fii_n:n\in\omega\}$ of intuitionistic tautologies of the form
\begin{equation}\label{eq:2}
\ET_{i<m}(p_i\lor p'_i)\to\Bigl(\ET_{j<m}(s_j\lor s'_j)\to\gamma(\vec p,\vec s,\vec{s'})\Bigr)
            \lor\Bigl(\ET_{k<m}(r_k\lor r'_k)\to\delta(\vec{p'},\vec r,\vec{r'})\Bigr),
\end{equation}
with $\gamma,\delta$ strict monotone, that have $\SF\IPC$ proofs constructible in time $n^{O(1)}$, but require
$\EF L$ proofs of size $2^{n^{\Omega(1)}}$ for any si logic~$L$ of unbounded branching.
\noproof\end{Thm}
($L$ has bounded branching if, roughly, there is a finite constant~$b$ such that any point in a finite $L$-frame has at
most $b$ immediate successors. See \cite{ej:sfef} for details.)

We would like to modify these tautologies further to make them purely implicational while preserving the bounds on the
lengths of their proofs. Rather than dwelling on these specific examples, we would like to have a method applicable
more generally to arbitrary tautologies; the ideal result could look as follows:
\begin{FThm}\label{fthm:elim-tau}
Given a formula~$\fii$, we can construct in polynomial time an implicational formula $\fii'$ with the following
property. If $P$ denotes $\Fr L$, $\EF L$, or $\SF L$ for a si logic~$L$, then $P$-proofs of $\fii$ and~$\fii'$ can be
constructed from each other in polynomial time.
\end{FThm}

As we will discuss in more detail below, this is too good to be true: we need to restrict the class of
tautologies or the class of logics. Nevertheless, we will seek results in a similar spirit.

We start with a construction that applies to all si logic in a very uniform way, but does not in general produce
fully implicational tautologies from arbitrary formulas~$\fii$. The idea is very simple: we introduce extension
variables for subformulas of~$\fii$. This cuts the meat of~$\fii$ into small chunks (extension axioms), most of which
can be rewritten with implicational formulas. (We stress that extension variables can be used in this way even in
Frege. The point is that here the set of extension axioms is a priori polynomially bounded and we can incorporate them in
the tautology, whereas an $\EF L$ proof may potentially use an unlimited number of extension axioms that cannot be
simulated in this way.)

We recall the usual notion of positive and negative occurrences of subformulas of~$\fii$: the occurrence of $\fii$ in
itself is positive, polarity is flipped when passing to the antecedent of an implication, and it is preserved by
other connectives.

\begin{Prop}\label{prop:impsimall}
Given a formula~$\fii$, we can construct in polynomial time a formula~$\ob\fii$ of the form
\[(p_0\to\bot)\to(p_1\to q_1\lor r_1)\to\dots\to(p_m\to q_m\lor r_m)\to\fii_\to,\]
where $\fii_\to$ is purely implicational, $m$ is the number of disjunctions with negative occurrences in~$\fii$, and the
$p_0\to\bot$ term is omitted if~$\fii$ has no negative occurrences of~$\bot$, with the following properties.
\begin{enumerate}
\item\label{item:2}
There are a poly-time constructible substitution~$\sigma$, and a poly-time constructible $\Fr\IPC$ proof of
$\sigma(\ob\fii)\to\fii$.
\item\label{item:1}
There is a poly-time constructible $\Fr\IPC$ proof of $\fii\to\ob\fii$.
\end{enumerate}
In particular, if $L$ is a si logic, and $P$ is $\Fr L$, $\EF L$, or $\SF L$, then $P$-proofs of $\fii$ and~$\ob\fii$ are
poly-time constructible from each other.
\end{Prop}
\begin{Pf}
We introduce a variable $p_\psi$ for each subformula $\psi\sset\fii$, where the original variables $p_i$ are identified
with~$p_{p_i}$. We define a set of formulas~$\Xi$, consisting of the following formulas for each nonvariable
$\psi\sset\fii$, depending on whether $\psi$~occurs positively or negatively in~$\fii$:
\begin{center}
\begin{tabular}{crl}
$\psi$&\multicolumn1c{positive}&\multicolumn1c{negative}\\
\hline
$\bot$&&$p_\psi\to\bot$\\[1\jot]
$\psi_0\to\psi_1$&$(p_{\psi_0}\to p_{\psi_1})\to p_\psi$&$p_\psi\to p_{\psi_0}\to p_{\psi_1}$\\[1\jot]
$\psi_0\land\psi_1$&$p_{\psi_0}\to p_{\psi_1}\to p_\psi$&$p_\psi\to p_{\psi_0}$\\
&&$p_\psi\to p_{\psi_1}$\\[1\jot]
$\psi_0\lor\psi_1$&$p_{\psi_0}\to p_\psi$&$p_\psi\to p_{\psi_0}\lor p_{\psi_1}$\\
&$p_{\psi_1}\to p_\psi$&
\end{tabular}
\end{center}
Enumerate $\Xi=\{\xi_0,\dots,\xi_r\}$ so that the formulas involving $\bot$ or~$\lor$ come first, and define $\ob\fii$
as
\[\xi_0\to\dots\to\xi_r\to p_\fii.\]

\ref{item:2}: Put $\sigma(p_\psi)=\psi$. Then the formulas from $\sigma(\Xi)$ are instances of constant-size
tautologies, hence they have linear-size proofs, using which it is easy to construct a proof of
$\sigma(\ob\fii)\to\sigma(p_\fii)$. 

\ref{item:1}: Let $\pi$ consist of $\Xi$ followed by formulas
\begin{equation}\label{eq:4}
\psi\to p_\psi
\end{equation}
for positively occurring $\psi\sset\fii$, and 
\begin{equation}\label{eq:5}
p_\psi\to\psi
\end{equation}
for negatively occurring $\psi\sset\fii$, ordered bottom-up. We can easily complete $\pi$ to a valid derivation of $\fii\to
p_\fii$ from assumptions~$\Xi$: for example, let $\psi=\psi_0\to\psi_1$ have positive occurrence in~$\fii$. Then
$\psi\to p_\psi$ has linear-size derivation from the formulas
\begin{gather*}
(p_{\psi_0}\to p_{\psi_1})\to p_\psi\\
p_{\psi_0}\to\psi_0\\
\psi_1\to p_{\psi_1}
\end{gather*}
that appear earlier in~$\pi$.

Using the feasible deduction theorem (and possibly Lemma~\ref{lem:struct} to reorder the formulas), we can turn $\pi$ into
a proof of $\fii\to\Xi\to p_\fii$, which is $\fii\to\ob\fii$.
\end{Pf}

This is not yet the end of the story. Let us see where Proposition~\ref{prop:impsimall} gets us with respect to the original
goal. We did not eliminate $\bot$ embedded in $\alpha,\beta$ in~\eqref{eq:1}, but this is not a problem, as
a $\bot$-free version of these tautologies is already provided by~\eqref{eq:2}. Elimination of $\lor$ is more
interesting. Most crucially, the main disjunctions on the right-hand sides of \eqref{eq:1}, \eqref{eq:2} occur
positively, and therefore \emph{are} eliminated by Proposition~\ref{prop:impsimall}. The same applies to $\lor$ embedded in
$\gamma,\delta$ of~\eqref{eq:2}. There might be some disjunctions coming from $\alpha,\beta$ left in~\eqref{eq:1},
however we can ignore these as we can apply classical reasoning inside negations. The remaining
disjunctions in \eqref{eq:1} and~\eqref{eq:2} occur in the context $\ET_i(p_i\lor p'_i)\to\cdots$, and we will get rid
of them by exploiting the intuitionistic equivalences
\begin{equation}\label{eq:19}
\left.
\begin{aligned}
(\fii\lor\psi\to\chi)&\eq(\fii\to\chi)\land(\psi\to\chi),\\
(\fii\land\psi\to\chi)&\eq(\fii\to\psi\to\chi),\\
(\fii\to\chi)&\eq(((\fii\to\chi)\to\chi)\to\chi).
\end{aligned}
\qquad\right\}
\end{equation}
We now formalize these considerations.
\begin{Def}\label{def:ess}
An occurrence of $\psi\lor\chi$ in a formula~$\fii$ is \emph{inessential} if
\begin{enumerate}
\item it is positive, or
\item it is in the scope of a negation (i.e., inside a subformula of~$\fii$ of the form $\alpha\to\bot$), or
\item there is an implication $\alpha\to\beta\sset\fii$ such that the given occurrence of $\psi\lor\chi$ is
inside~$\alpha$, and not inside any subformula of~$\alpha$ that is an implication. In other words, the only connectives
on the path from $\psi\lor\chi$ to~$\alpha$ in the formula tree are $\lor$ and~$\land$.
\end{enumerate}
All other occurrences of disjunctions in~$\fii$ are \emph{essential}.
\end{Def}
\begin{Thm}\label{prop:impessall}
Proposition~\ref{prop:impsimall} holds with $m$ being the number of essential occurrences of disjunctions in~$\fii$.
\end{Thm}
\begin{Pf}
For any subformula $\alpha\to\bot$ of~$\fii$, we may replace $\alpha$ with a canonically chosen classically equivalent
formula using $\to,\bot$ as the only connectives, hence we may assume without loss of generality that no disjunctions
occur inside negations.

We modify the construction from the proof of Proposition~\ref{prop:impsimall} as follows. Consider any positively occurring
implication $\gamma\sset\fii$, and write it as
\[\mu(\alpha_0,\dots,\alpha_{r-1})\to\beta,\]
where $\mu(x_0,\dots,x_{r-1})$ is a strict monotone formula, and each $\alpha_i$ is a variable, a constant, or an implication. We omit the
variables $p_{\nu(\vec\alpha)}$ for non-variable subformulas $\nu\sset\mu$ along with the associated formulas in~$\Xi$,
and introduce instead variables $q_\nu^\gamma$ for~$\nu\sset\mu$ with intended meaning $\nu(\vec\alpha)\to\beta$, where
$q^\gamma_\mu$ is identified with~$p_\gamma$. We will write just~$q_\nu$ to simplify the notation.

We include in~$\Xi$ the following formulas for each $\nu\sset\mu$:
\begin{center}
\begin{tabular}{cr}
$\nu$&\multicolumn1c{formula in $\Xi$}\\
\hline
$x_i$&$(p_{\alpha_i}\to p_\beta)\to q_\nu$\\
$\nu'\lor\nu''$&$((q_{\nu'}\to q_{\nu''}\to p_\beta)\to p_\beta)\to q_\nu$\\
$\nu'\land\nu''$&$((q_{\nu'}\to p_\beta)\to q_{\nu''})\to q_\nu$
\end{tabular}
\end{center}

\ref{item:2}: We extend $\sigma$ by putting $\sigma(q_\nu)=(\nu(\vec\alpha)\to\beta)$. As in the proof of
Proposition~\ref{prop:impsimall}, it is enough to construct short $\Fr\IPC$ proofs of the new formulas in~$\sigma(\Xi)$.
This is straightforward:
\begin{align*}
(\nu'(\vec\alpha)\lor\nu''(\vec\alpha)\to\beta)
&\eq(((\nu'(\vec\alpha)\lor\nu''(\vec\alpha)\to\beta)\to\beta)\to\beta)\\
&\eq(((\nu'(\vec\alpha)\to\beta)\land(\nu''(\vec\alpha)\to\beta)\to\beta)\to\beta)\\
&\eq(((\nu'(\vec\alpha)\to\beta)\to(\nu''(\vec\alpha)\to\beta)\to\beta)\to\beta)
\end{align*}
using~\eqref{eq:19}; likewise,
\begin{align*}
(\nu'(\vec\alpha)\land\nu''(\vec\alpha)\to\beta)
&\eq(\nu''(\vec\alpha)\to\nu'(\vec\alpha)\to\beta)\\
&\eq(\nu''(\vec\alpha)\to((\nu'(\vec\alpha)\to\beta)\to\beta)\to\beta)\\
&\eq(((\nu'(\vec\alpha)\to\beta)\to\beta)\to\nu''(\vec\alpha)\to\beta).
\end{align*}

\ref{item:1}: We modify the proof~$\pi$ considered in Proposition~\ref{prop:impsimall} by including the
formulas
\begin{equation}\label{eq:3}
(\nu(\vec\alpha)\to\beta)\to q_\nu
\end{equation}
for all $\nu\sset\mu$. Notice that the final formula \eqref{eq:3} for $\nu=\mu$ is just
\eqref{eq:4} for $\psi=\gamma$. We complete $\pi$ to a valid derivation as in the proof of Proposition~\ref{prop:impsimall}; we
only need to take care of the new formulas~\eqref{eq:3}.

If $\nu=\nu'\lor\nu''$, we can derive
\begin{align*}
(\nu(\vec\alpha)\to\beta)&\to(\nu'(\vec\alpha)\lor\nu''(\vec\alpha)\to p_\beta)
      &&\text{\eqref{eq:4} for $\beta$}\\
&\to(((\nu'(\vec\alpha)\to p_\beta)\to(\nu''(\vec\alpha)\to p_\beta)\to p_\beta)\to p_\beta)
      &&\text{\eqref{eq:19}}\\
&\to((q_{\nu'}\to q_{\nu''}\to p_\beta)\to p_\beta)
      &&\text{\eqref{eq:3} for $\nu',\nu''$}\\
&\to q_\nu,
      &&\Xi
\end{align*}
using the fact that $q_{\nu'}$ and $q_{\nu''}$ occur positively in $(q_{\nu'}\to q_{\nu''}\to p_\beta)\to p_\beta$.

The other cases are analogous.
\end{Pf}
\begin{Cor}\label{cor:expsep}
There are implicational formulas $\{\fii_n:n\in\omega\}$ satisfying Theorem~\ref{thm:sfef-lb}.
\end{Cor}
\begin{Pf}
No essential disjunctions occur in~\eqref{eq:2}.
\end{Pf}
(See Theorem~\ref{thm:sep-imp} for a further improvement of Corollary~\ref{cor:expsep}.)

As we will see, we can strengthen Proposition~\ref{prop:impsimall} for intuitionistic logic so that the resulting formula
is always purely implicational, irrespective of what kind of disjunctions occur in~$\fii$. However, this will require a
construction of a different kind that no longer applies to arbitrary si logics~$L$. The reason is that the properties
of~$\ob\fii$ as given by Proposition~\ref{prop:impsimall} imply that $\ob\fii$ is interderivable with~$\fii$ in the sense
that $\IPC+\fii=\IPC+\ob\fii$. If we can always find a positive (i.e., $\bot$-free) $\ob\fii$ with this property for
every $L$-tautology $\fii$, then $L$ is positively axiomatizable over~$\IPC$, or equivalently, the class of
general $L$-frames is closed under dense subframes. For instance, this fails for
$L=\lgc{KC}$, which has the same positive fragment as~$\IPC$. Similarly, if we can always find a $\lor$-free $\ob\fii$, then
$L$  is a cofinal-subframe logic.

This problem comes up even if we do not ask $\ob\fii$ to have the strong properties guaranteed by
Proposition~\ref{prop:impsimall}, but merely that it is positive, and preserves proof length in a particular logic~$L$ up to a
polynomial. If $L\sset\lgc{KC}$, we know from \cite[L.~6.30]{ej:sfef} that $\EF L$ is polynomially equivalent
to~$\EF\IPC$ with respect to positive tautologies. Thus, the existence of such a map $\fii\mapsto\ob\fii$ would imply
that $\EF L$ has essentially the same proof complexity as $\EF\IPC$, and in particular, that every $L$-tautology has a
proof of at most exponential size. However, there are a vast number of logics between
$\IPC$ and~$\lgc{KC}$, many of which are outside~$\mathrm{NEXP}$, or even undecidable.%
\footnote{The undecidable, finitely axiomatized si logics with the disjunction property (which incidentally implies
they also have the same $\lor$-free fragment as~$\IPC$) from \cite[Ex.~15.11]{cha-zax} are easily seen to be included
in~$\lgc{KC}$. As for natural examples, the author wonders what is the complexity of the Kreisel--Putnam logic.}

So, to make a long story short, we will only show how to eliminate $\bot$ and~$\lor$ from tautologies in logics whose
proper axioms do not contain the respective connectives, which appears to be the best we can hope for. We start with the
simpler case of~$\bot$. The idea here is just that the conjunction of all variables is a good enough approximation
of~$\bot$ vis-\`a-vis positive formulas.
\begin{Thm}\label{prop:bot-free}
Given a formula~$\fii$, we can construct in polynomial time a positive formula~$\fii^+$ and a substitution~$\sigma$
with the following properties.
\begin{enumerate}
\item\label{item:4}
There is a poly-time constructible $\Fr\IPC$ proof of $\sigma(\fii^+)\to\fii$.
\item\label{item:3}
Let $P$ denote $\Fr L$, $\EF L$, or $\SF L$ for a si logic~$L$ axiomatizable by positive formulas. Then given a
$P$-proof of~$\fii$, we can construct in polynomial time a $P$-proof of $\fii^+$.
\end{enumerate}
Moreover, we can make $\fii^+$ to have the form
\begin{equation}\label{eq:10}
(p_1\to q_1\lor r_1)\to\dots\to(p_m\to q_m\lor r_m)\to\fii_\to,
\end{equation}
where $\fii_\to$ is purely implicational, and $m$ is the number of essential occurrences of $\lor$ in~$\fii$.
\end{Thm}
\begin{Pf}
Write $\fii=\fii'(p_0,\dots,p_{n-1},\bot)$, where $\fii'$ is positive, and $p_0,\dots,p_{n-1}$ are all variables occurring
in~$\fii$. Define $\fii^+$ as
\begin{equation}\label{eq:8}
(r\to p_0)\to\dots\to(r\to p_{n-1})\to\fii'(p_0,\dots,p_{n-1},r),
\end{equation}
where $r$ is a new variable. \ref{item:4} is straightforward, using the substitution $\sigma(r)=\bot$.

\ref{item:3}: We can assume the Frege rules of~$P$ consist of positive axioms, modus ponens, and the
axiom~\eqref{eq:7}.
First, let $P$ be $\Fr L$ or $\EF L$. Let $\pi$ be a $P$-proof of~$\fii$. We may
assume that the only variables occurring in~$\pi$ are $p_i$ and extension variables, and that none of the extension
variables is~$r$. We can also assume that the only instances of~\eqref{eq:7} in~$\pi$ are with $\gamma=p_i$: we can
derive $\bot\to\gamma$ for an arbitrary~$\gamma$ from these by a polynomial-size subproof by induction on the
complexity of~$\gamma$ (using extension axioms to get $\bot\to q_j$ for extension variables~$q_j$).

We replace $\bot$ with~$r$ in the proof. Treating the axioms $r\to p_i$ as extra assumptions, we construct a
proof of~$\fii^+$ by an application of the feasible deduction theorem.

Now, let $P=\SF L$, and $\pi=\fii_1,\dots,\fii_m$ be a $P$-proof of $\fii_m=\fii$. We may assume $r$ does not occur
in~$\pi$. We will complete $\fii_1^+,\dots,\fii_m^+$ to a valid $P$-proof, where all the formulas $\fii_j^+$ use the
same~$r$, but only variables actually occurring in~$\fii_j$ in the premise of~\eqref{eq:8}.

$\SF L$ has a constant finite set of axioms, whose translations have constant-size proofs, which takes care of the case
when $\fii_j$ is an axiom.

If $\fii_j$ is derived by modus ponens from $\fii_k$ and $\fii_l=(\fii_k\to\fii_j)$,
let $p_0,\dots,p_{n-1}$ be the variables of $\fii_l$. We can insert extra conjuncts in~$\fii_k^+$ if necessary to
obtain
\begin{align*}
\{r\to p_i:i<n\}&\to\fii_k'(\vec p,r),\\
\{r\to p_i:i<n\}&\to\fii_k'(\vec p,r)\to\fii'_j(\vec p,r),\\
\intertext{from which we derive}
\{r\to p_i:i<n\}&\to\fii_j'(\vec p,r)
\end{align*}
using Lemma~\ref{lem:struct}. We get rid of unwanted conjuncts $r\to p_i$ for $p_i$ not occurring in~$\fii_j$ by
substituting $r$ for~$p_i$.

Assume that $\fii_j(p_0,\dots,p_{n-1})$ is derived from~$\fii_k(q_0,\dots,q_{m-1})$ by the substitution
rule, so that $\fii_j=\fii_k(\alpha_0,\dots,\alpha_{m-1})$ for some formulas $\alpha_i(\vec p)$, $i<m$. (Here, $\vec q$
may well be the same variables as~$\vec p$; we distinguish them in notation for clarity.)
We are given the formula~$\fii_k^+$, which is
\[\{r\to q_l:l<m\}\to\fii_k'(\vec q,r).\]
Applying the substitution rule to~$\fii_k^+$, we derive
\[\{r\to\alpha'_l(\vec p,r):l<m\}\to\underbrace{\fii'_k(\alpha'_0(\vec p,r),\dots,\alpha'_{m-1}(\vec p,r),r)}_{\fii'_j(\vec p,r)}.\]
We construct short proofs of
\[\{r\to p_i:i<n\}\to r\to\alpha'_l(\vec p,r)\]
for $l<m$ by induction on the complexity of the positive formulas~$\alpha'_l(\vec p,r)$, and conclude
\[\{r\to p_i:i<n\}\to\fii'_j(\vec p,r),\]
which is~$\fii_j^+$. This finishes the argument for $P=\SF L$.

Finally, notice that the construction of~$\fii^+$ did not increase the number of essential (or inessential, for
that matter) disjunctions, hence if we use $\ob{\fii^+}$ from Theorem~\ref{prop:impessall} instead, it will have the
required form~\eqref{eq:10}.
\end{Pf}

Elimination of~$\lor$ is ostensibly more complicated than elimination of~$\bot$, but it can be seen as a generalization
of the latter, as $\bot$ is just the empty disjunction. In the proof of Theorem~\ref{prop:bot-free}, we exploited the fact
that the algebra of positive formulas in finitely many variables has a least element (and therefore is a Heyting
algebra), which we can use to emulate~$\bot$. The least element should be the meet of all positive formulas, but this
equals the meet of just variables, which makes it definable by a finite formula.

In order to eliminate~$\lor$ in a similar way, Diego's theorem \cite[\S5.4]{cha-zax} ensures that the algebra of $\lor$-free formulas in
finitely many variables is finite; since it is a $\land$-semilattice, it is in fact a complete lattice (and a Heyting
algebra). The join operation of this lattice can be used to emulate disjunction. The join of $\psi$ and $\chi$ should be the
conjunction of all $\lor$-free formulas implied by $\psi\lor\chi$, which are up to equivalence exactly the formulas of
the form
\[(\psi\to\gamma)\to(\chi\to\gamma)\to\gamma.\]
As with~$\bot$, it turns out that it is enough to consider just variables (and~$\bot$) in place of~$\gamma$, resulting
in a short conjunction.

We will need to emulate all disjunctions appearing in the proof. Since the replacement
for~$\psi\lor\chi$ involves many occurrences of $\psi$ and~$\chi$, we need to work with circuits to preserve polynomial
size. That is, the argument will only give polynomial simulation of $\M{EF}$ systems, not Frege.
\begin{Thm}\label{prop:lor-free}
Given a formula~$\fii$, we can construct in polynomial time a $\{\to,\bot\}$-formula~$\tilde\fii$ (purely implicational
if $\fii$ is positive) and a substitution~$\sigma$ with the following properties.
\begin{enumerate}
\item\label{item:5}
There is a poly-time constructible $\Fr\IPC$ proof of $\sigma(\tilde\fii)\to\fii$.
\item\label{item:6}
Let $P$ denote $\EF L$ or $\SF L$ for a cofinal-subframe si logic~$L$. Given a
$P$-proof of~$\fii$, we can construct in polynomial time a $P$-proof of $\tilde\fii$.
\end{enumerate}
\end{Thm}
\begin{Pf}
We will first construct a $\lor$-free formula~$\tilde\fii$ with these properties.

Let a formula $\fii(\vec p)$ be given. For each $\psi\lor\chi\sset\fii$, we introduce a new variable
$r_{\psi\lor\chi}$; we denote by $\alpha^*$ the result of replacing all top-most occurrences of disjunctions
$\psi\lor\chi$ with~$r_{\psi\lor\chi}$. We define
\[\tilde\fii=\Delta_\fii\to\fii^*,\]
where the set~$\Delta_\fii$ consists of the formulas
\begin{gather}
\label{eq:9}\psi^*\to r_{\psi\lor\chi},\\
\label{eq:16}\chi^*\to r_{\psi\lor\chi},\\
\label{eq:17}r_{\psi\lor\chi}\to(\psi^*\to v)\to(\chi^*\to v)\to v
\end{gather}
for $\psi\lor\chi\sset\fii$, and $v\in V_\fii=\{\vec p,\vec r,\bot\}$ (where $\vec r$ stands for the list $\{r_{\psi\lor\chi}:\psi\lor\chi\sset\fii\}$).

\ref{item:5}: Putting $\sigma(r_{\psi\lor\chi})=\psi\lor\chi$, we have $\sigma(\alpha^*)=\alpha$ for each~$\alpha$,
hence $\sigma(\tilde\fii)=\sigma(\Delta_\fii)\to\fii$, and the formulas in~$\sigma(\Delta_\fii)$ are instances of
constant-size tautologies.

\ref{item:6}: First, let $P=\EF L$. We may assume the Frege rules of~$P$ are instances of $\lor$-free axioms, modus
ponens, and the axioms~\eqref{eq:11}--\eqref{eq:13}.
Let $\pi$ be a $P$-proof of~$\fii$. We may assume the only variables in~$\pi$ are $\vec p$ and extension variables. We
introduce new variables $r_{\psi\lor\chi}$ for each disjunction appearing in~$\pi$, and write $\alpha^*$ for the result
of replacing all top-most disjunctions in~$\alpha$ with the corresponding variables. Notice that this agrees with the
definition of~$\alpha^*$ used in the construction of~$\tilde\fii$.

We replace each formula $\alpha$ in~$\pi$ with~$\alpha^*$. At the beginning of the proof, we add the
formulas~$\Delta_\fii$ as extra assumptions. We introduce the extension axiom
\begin{equation}\label{eq:14}
r_{\psi\lor\chi}\eq\ET_{v\in V_\fii}\bigl((\psi^*\to v)\to(\chi^*\to v)\to v\bigr)
\end{equation}
for each $\psi\lor\chi\nsset\fii$; we place this axiom after extension axioms for all extension variables occurring
in $\psi^*,\chi^*$. We stress that $V_\fii$ does \emph{not} contain variables~$r_{\psi\lor\chi}$ for $\psi\lor\chi\nsset\fii$.

In order to make this a valid derivation, we need to fix occurrences of \eqref{eq:11}--\eqref{eq:13} in the proof. As
for \eqref{eq:11} and~\eqref{eq:12}, the formulas
\[\alpha^*\to r_{\alpha\lor\beta},\qquad\beta^*\to r_{\alpha\lor\beta}\]
are included in~$\Delta_\fii$ for $\alpha\lor\beta\sset\fii$, and easily provable from~\eqref{eq:14}
for other $\alpha\lor\beta$.

As for~\eqref{eq:13}, notice that $\gamma^*$ is $\lor$-free. The formulas
\begin{equation}\label{eq:15}
r_{\alpha\lor\beta}\to(\alpha^*\to\xi)\to(\beta^*\to\xi)\to\xi
\end{equation}
are in~$\Delta_\fii$ or derivable from~\eqref{eq:14} when $\xi\in V_\fii$. For general $\lor$-free $\xi$ (using
only variables occurring in the proof), we construct a derivation of~\eqref{eq:15} by inner induction on the complexity of~$\xi$, and outer induction
on the extension variables (old or new) occurring in~$\xi$. For the induction hypothesis, if $\circ\in\{\to,\land\}$,
then
\begin{align*}
&r_{\alpha\lor\beta}\to(\alpha^*\to\xi_0)\to(\beta^*\to\xi_0)\to\xi_0,\\
&r_{\alpha\lor\beta}\to(\alpha^*\to\xi_1)\to(\beta^*\to\xi_1)\to\xi_1
\vdash r_{\alpha\lor\beta}\to(\alpha^*\to\xi_0\circ\xi_1)\to(\beta^*\to\xi_0\circ\xi_1)\to\xi_0\circ\xi_1
\end{align*}
is an instance of a constant-size derivation. If $\xi$ is an extension variable, we derive~\eqref{eq:15} for its
definition by the induction hypothesis, and use the extension axiom.

In the end, we obtain a valid $P$-derivation of~$\fii^*$ from assumptions~$\Delta_\fii$, and we use the feasible
deduction theorem to get a proof of $\tilde\fii$.

Now, let $P=\SF L$, and $\fii_1,\dots,\fii_m$ be a $P$-proof of~$\fii$. We consider the sequence
$\tilde\fii_1,\dots,\tilde\fii_m$, and make it a valid $P$-proof as follows.

As in the proof of Theorem~\ref{prop:bot-free}, translations of axioms have constant-size proofs.

If $\fii_j=\fii_k(\alpha_1,\dots,\alpha_n)$ is inferred from~$\fii_k$ by substitution, we rename variables
$r_{\psi\lor\chi}$ in~$\tilde\fii_k$ to $r_{\psi(\vec\alpha)\lor\chi(\vec\alpha)}$, and substitute
$\alpha_1^*,\dots,\alpha_n^*$ to obtain $\Delta'\to\fii_j^*$, where $\Delta'$ contains some of the formulas of the form
\begin{gather*}
\psi^*\to r_{\psi\lor\chi},\\
\chi^*\to r_{\psi\lor\chi},\\
r_{\psi\lor\chi}\to(\psi^*\to v)\to(\chi^*\to v)\to v
\end{gather*}
for $\psi\lor\chi\sset\fii_j$, and $v\in\{\alpha_1^*,\dots,\alpha_n^*,\vec r,\bot\}$. Only the last are a problem; we
construct their short derivations from~$\Delta_{\fii_j}$ by induction on the complexity of~$\alpha_i^*$ as in the proof
of~\eqref{eq:15}. Thus, we obtain a proof of~$\Delta_{\fii_j}\to\fii_j^*$, i.e., $\tilde\fii_j$.

If $\fii_j$ is derived
from $\fii_k$ and $\fii_l=(\fii_k\to\fii_j)$ by modus ponens, we insert
dummy premises in $\tilde\fii_k$ to obtain $\Delta_{\fii_l}\to\fii_k^*$, and infer
\[\Delta_{\fii_l}\to\fii_j^*.\]
We need to get rid of the unwanted conjuncts from~$\Delta_{\fii_l}\bez\Delta_{\fii_j}$. First, assume
$r_{\psi\lor\chi}$ occurs in~$\Delta_{\fii_l}$ for some $\psi\lor\chi\nsset\fii_j$, and take $\psi\lor\chi$ maximal such. We
substitute $r_{\psi\lor\chi}$ with the formula
\[\xi=\ET_{v\in V}\bigl((\psi^*\to v)\to(\chi^*\to v)\to v\bigr),\]
where $V$ is the set of~$v$ for which~\eqref{eq:17} occurs in the formula (minus $v=r_{\psi\lor\chi}$ itself, for which
the conjunct is redundant). The formulas \eqref{eq:9}--\eqref{eq:17} associated with~$r_{\psi\lor\chi}$ have short
proofs after the substitution, and we can delete them. We can also delete the formulas
\[r_{\psi'\lor\chi'}\to(\psi'^*\to\xi)\to(\chi'^*\to\xi)\to\xi\]
for other $r_{\psi'\lor\chi'}$: they have short derivations from the remaining formulas, as $\xi$ is a $\lor$-free
formula not containing~$r_{\psi\lor\chi}$. After that, there are no traces left of either $r_{\psi\lor\chi}$ or~$\xi$.

We continue in this way to eliminate all variables $r_{\psi\lor\chi}$ for $\psi\lor\chi\nsset\fii_j$. The only unwanted
things left are then formulas~\eqref{eq:17} for $\psi\lor\chi\sset\fii_j$, and $v=p_i$ a variable not occurring
in~$\fii_j$. We can e.g.\ substitute $\bot$ for $p_i$.

This concludes the proof of~\ref{item:6}, but we have so far only constructed a $\lor$-free formula~$\tilde\fii$ with
the required properties. In order to make it a $\{\to,\bot\}$-formula, it suffices to apply Proposition~\ref{prop:impsimall}.
However, we need to be a bit more careful if we want an implicational formula in case $\fii$ is positive, as the
definition of~$\tilde\fii$ unconditionally introduced $\bot$.

Fortunately, if $\fii$ is positive, the formulas~\eqref{eq:17} with $v=\bot$ are redundant in~$\Delta_\fii$: the remaining
formulas imply
\[r_{\psi\lor\chi}\to(\psi^*\to\bot)\to(\chi^*\to\bot)\to v,\qquad v\in\{\vec p,\vec r\}.\]
Since $\chi^*$ is positive, this yields
\[r_{\psi\lor\chi}\to(\psi^*\to\bot)\to(\chi^*\to\bot)\to\chi^*,\]
hence
\[r_{\psi\lor\chi}\to(\psi^*\to\bot)\to(\chi^*\to\bot)\to\bot\]
as needed.
\end{Pf}

For~$\IPC$, and more generally logics axiomatizable by both $\bot$-free and $\lor$-free formulas (i.e., subframe
logics), we can apply Theorems \ref{prop:bot-free} and~\ref{prop:lor-free} in a row:
\begin{Cor}\label{cor:bot-lor-free}
Given a formula~$\fii$, we can construct in polynomial time a purely implicational formula~$\hat\fii$ and a substitution~$\sigma$
with the following properties.
\begin{enumerate}
\item
There is a poly-time constructible $\Fr\IPC$ proof of $\sigma(\hat\fii)\to\fii$.
\item
Let $P$ denote $\EF L$ or $\SF L$ for a subframe si logic~$L$. Given a
$P$-proof of~$\fii$, we can construct in polynomial time a $P$-proof of $\hat\fii$.\noproof
\end{enumerate}
\end{Cor}

\section{Elimination of connectives from proofs}\label{sec:elim-proof}

So far we were concerned with elimination of connectives from the formula to be proved, but we still allowed proofs to
be carried out in the full logic. It is natural to ask if we can also eliminate connectives other than implication from
the proofs themselves, and indeed some of the arguments we used earlier can help with that goal, too.
The ideal result could be formulated as
\begin{FThm}\label{fthm:elim-pf}
Let $P$ denote $\Fr L$, $\EF L$, or $\SF L$ for a si logic~$L$, and ${\to}\in C\sset C_\IPC$. Given a $P$-proof of a
$C$-formula~$\fii$, we can construct in polynomial time a $P$-proof of~$\fii$ consisting only of $C$-formulas.
\end{FThm}
In this formulation, the result is extremely sensitive to the choice of axioms of~$L$: to begin with, there is no way
it could hold if we axiomatize $L$ so that none of the axioms is a $C$-formula. Thus, we can only get results of this
form for very particular axiom systems.

We could avoid this dependence on axiomatization by shifting perspective: we can view it as a feasible conservativity
result, i.e., a form of p-simulation between two different systems. For example, here is such a statement for~$\M{EF}$:
\begin{FThm}\label{fthm:elim-pf-cons}
Let $L$ be a si logic, and ${\to}\in C\sset C_\IPC$. Given an $\EF L$ proof of a $C$-formula~$\fii$, we can construct
in polynomial time an $\EF{L_C}$ proof of~$\fii$. That is, $\EF{L_C}$ p-simulates $\EF L$ with respect to $C$-formulas.
\end{FThm}

Nevertheless, we start with a result in the form of~\ref{fthm:elim-pf}; in fact, we formulate it in a stronger way,
which gives us control over unwanted connectives even if they appear in the tautology being proved. In a pleasant turn
of events, we have no real work to do: we just reap what we have grown in the proofs of
Theorems \ref{prop:bot-free} and~\ref{prop:lor-free}.

For simplicity, we state the result only for extended Frege. The less interesting
part~\ref{item:7} also works for Frege. The statement also holds for substitution Frege after appropriate modifications
($\M{SF}$ proofs only include the basic form of each axiom once, and if needed, its instances can be derived by means
of the substitution rule, thus the restrictions on instances below are not meaningful).
\begin{Thm}\label{prop:lor-bot-pfs}
Let $L$ be a si logic, and $P$ a standard $\EF L$ system. Given a $P$-proof $\pi$ of a formula~$\fii$, we can construct in polynomial
time a $P$-proof $\pi'$ of~$\fii$ with the following properties:
\begin{enumerate}
\item\label{item:8}
If the proper axiom of~$P$ is $\lor$-free, then the only disjunctions occurring in~$\pi'$ are subformulas
of~$\fii$, and the only instances of~\eqref{eq:13} in~$\pi'$ are with $\gamma$ a subformula of~$\fii$ which is a
variable, a disjunction, or~$\bot$.
\item\label{item:7}
If $\fii$ and the proper axiom of~$P$ are positive, then $\bot$ does not occur in~$\pi'$.
Otherwise the only instances of~\eqref{eq:7} in~$\pi'$ are with $\gamma$ a variable from~$\fii$.
\end{enumerate}
\end{Thm}
\begin{Pf}
\ref{item:8}: Let $\tilde\fii$ be the $\lor$-free (and $\bot$-free, if $\fii$ is) formula from Theorem~\ref{prop:lor-free},
and $\tilde\pi$ its $P$-proof constructed there. There are no disjunctions in~$\tilde\pi$. We apply to~$\tilde\pi$
the substitution $\sigma(r_{\psi\lor\chi})=\psi\lor\chi$. Then the only disjunctions in~$\sigma(\tilde\pi)$ are
subformulas of~$\fii$, and the proof does not use axioms \eqref{eq:11}--\eqref{eq:13}. The conclusion of
the proof is $\sigma(\Delta_\fii)\to\fii$, and the formulas in $\sigma(\Delta_\fii)$ are instances of
\eqref{eq:11}--\eqref{eq:13} of the required form, hence we can include them in the proof, and infer~$\fii$ by modus
ponens.

The argument for~\ref{item:7} is analogous: we take the $\bot$-free proof of the formula~$\fii^+$ constructed in
Theorem~\ref{prop:bot-free}. If $\fii$ contains~$\bot$, we substitute $\bot$ for~$r$, and eliminate the premises $\bot\to
p_i$ by introducing them as axioms. If $\fii$ is positive, we substitute $\ET_ip_i$ for~$r$ instead. This construction
does not introduce any new disjunctions or instances of~\eqref{eq:13}, hence if the axioms of~$L$ are $\bot$-free
and $\lor$-free, we can satisfy \ref{item:8} and~\ref{item:7} simultaneously.
\end{Pf}

Theorem~\ref{prop:lor-bot-pfs} quite satisfactorily eliminates $\lor$ and~$\bot$. Unfortunately, not only it does not
eliminate~$\land$, the argument may even introduce new conjunctions into the proof (cf.~\eqref{eq:14}). We will spend
the rest of this section by trying to remedy this defect in several ways.

Let us first note that while every si logic can be axiomatized without the use of~$\land$ by Lemma~\ref{lem:conj-free}, we
cannot simply extend Theorem~\ref{prop:lor-bot-pfs} by adding a conjunction clause analogous to~\ref{item:8}. Let $\alpha$
be an implicational axiom. Completely disregarding lengths of proofs, Theorem~\ref{prop:lor-bot-pfs} implies that the
$\lor$-free fragment of the logic $L=\IPC+\alpha$ can be axiomatized by $\alpha$ over the $\lor$-free fragment
of~$\IPC$, and similarly for~$\bot$. As shown in Proposition~\ref{prop:frag}, similar statements are not true in the absence
of~$\land$: for example, it may happen that $\IPC_\to+\alpha$ is strictly weaker than $L_\to$.
This does not mean that elimination of connectives from proofs besides the cases in
Theorem~\ref{prop:lor-bot-pfs} is impossible, but it cannot work so uniformly; we will need to impose extra requirements on
the logics involved, and on their axiom sets.

We include the following particular results. Section~\ref{sec:elim-bot} contains an alternative simple
transformation for elimination of~$\bot$ that does not rely on the presence of other connectives such as~$\land$, and
works even for Frege, but needs peculiar assumptions on the logic. In Section~\ref{sec:elim-land} we present a method
for elimination of~$\land$ from $\lor$-free proofs for axiom systems that avoid the obstacle from Proposition~\ref{prop:frag}.
To tie a loose end, we give an ad~hoc argument that the implicational tautologies from Corollary~\ref{cor:expsep} have short
$\SF{\IPC_\to}$ proofs in Section~\ref{sec:expon-separ}.

\subsection{Light-weight elimination of $\bot$}\label{sec:elim-bot}

We are going to present a method for elimination of $\bot$ from intuitionistic proofs. It is less invasive than the one from
Theorem~\ref{prop:lor-bot-pfs}, and in particular, it avoids introducing $\land$ (or other 
connectives, for that matter) into the proof. The translation also applies to other logics, under a certain semantic
condition: the class of frames for the logic must be preserved under attachment of a new top element. (Thus, it does
not fully supersede Theorem~\ref{prop:lor-bot-pfs} \ref{item:7}.) Interestingly,
unlike other simulations in this paper, this one can even be usefully employed to translate proofs from one logic into
another; in fact, its variant was introduced in \cite[L.~6.30]{ej:sfef} for that very purpose. Another presentation of
this translation, as well as the associated semantic construction in Definition~\ref{def:model-top}, was later given by
De~Jongh and Zhao~\cite{dj-zz:pos}.

\begin{Def}\label{def:model-top}
If $\cF=\p{F,\le}$ is a Kripke frame, let $\cF^t=\p{F^t,\le^t}$ denote the frame obtained from~$F$ by attaching a new
top element~$t$; that is, $F^t=F\cup\{t\}$, and ${\le^t}={\le}\cup(F^t\times\{t\})$. We can also
apply the construction to general frames $\cF=\p{F,\le,A}$: we define $\cF^t=\p{F^t,\le^t,A^t}$ with the same $F^t$
and~$\le^t$ as before, and
$A^t=\{\nul\}\cup\bigl\{X\cup\{t\}: X\in A\bigr\}$.
\end{Def}
\begin{Thm}\label{thm:bot-top}
Given an $\Fr\IPC$ proof $\pi$ of a positive formula~$\fii$, we can construct in polynomial time an $\Fr{\IPC}$ proof
of~$\fii$ containing no~$\bot$, and only those positive connectives that already occur in~$\pi$.

More generally, assume that $L^1$ is a si logic complete wrt a class of frames~$\mathcal K$, and $L^0$ is a si logic
which is valid in~$\cF^t$ for all $\cF\in\mathcal K$. Let $C\sset\{{\to},{\land},{\lor}\}$.
Given an $\Fr{L^0_{C,\bot}}$ proof of a $C$-formula~$\fii$, we can construct in
polynomial time an $\Fr{L^1_C}$ proof of~$\fii$.

The same results hold with $\M{EF}$ in place of Frege.
\end{Thm}
\begin{Pf}
Let $v$ be the classical assignment that makes all variables true. For every $C\cup\{\bot\}$-formula~$\fii$ such that
$v(\fii)=1$, we define\footnote{We leave $\fii^*$ undefined when $v(\fii)=0$. If we needed it for other purposes, the
semantically correct way to define it would be $\fii^*=\bot$, which is however not a positive formula.}
a $C$-formula $\fii^*$ by induction on the complexity of~$\fii$:
\begin{align*}
p^*&=p,\\
(\fii\land\psi)^*&=\fii^*\land\psi^*,\\
(\fii\lor\psi)^*&=\begin{cases}
  \fii^*\lor\psi^*,&v(\fii)=v(\psi)=1,\\
  \fii^*,&v(\psi)=0,\\
  \psi^*,&v(\fii)=0,
\end{cases}\\
(\fii\to\psi)^*&=\begin{cases}
  \fii^*\to\psi^*,&v(\fii)=v(\psi)=1,\\
  \top,&v(\fii)=0.
\end{cases}
\end{align*}
Notice that if $\fii$ contains no~$\bot$, then $v$ satisfies all its subformulas, hence $\fii^*=\fii$.

If $L^1$ is inconsistent, every $C$-formula has a linear-size $\Fr{L^1_C}$ proof. Thus, we can assume $L^1$ is
consistent, in which case so is $L^0$ (as $\mathcal K\ne\nul$), hence $L^0\sset\CPC$.

Let $\pi=\p{\fii_0,\dots,\fii_m}$ be an $\Fr{L^0_{C,\bot}}$ proof of a $C$-formula $\fii_m$. By our assumption, all
$\fii_i$ are classical tautologies, in particular $v(\fii_i)=1$, hence we can construct the sequence
$\pi^*=\p{\fii_0^*,\dots,\fii_m^*}$. Clearly, $\pi^*$ consists of $C$-formulas, and ends with $\fii_m^*=\fii_m$. We need
to show that it can be completed to a polynomially larger $\Fr{L^1_C}$ proof.
\begin{Cl}\label{cl:l1taut}
For any $L^0$-tautology $\fii$, the formula~$\fii^*$ is an $L^1$-tautology.
\end{Cl}
\begin{Pf*}
It suffices to show that if $\cF\in\mathcal K$ and $\model$ is an admissible valuation in~$\cF$, then $\cF\model\fii^*$.
Let $\model^t$ be the valuation in~$\cF^t$ which agrees with~$\model$ on variables in points of~$F$, and makes all
variables true in the new top point~$t$. It is easy to show by induction on the complexity of a formula~$\alpha$ that
\begin{align*}
\cF^t,t\model\alpha&\iff v(\alpha)=1,\\
\cF^t,x\model\alpha&\iff v(\alpha)=1\text{ and }\cF,x\model\alpha^*
\end{align*}
for $x\in F$. Since $\cF^t$ is an $L^0$-frame, it follows that $\cF,x\model\fii^*$ for all $x\in F$.
\end{Pf*}

Claim~\ref{cl:l1taut} shows that instances of axioms in~$\pi$ translate to $L^1_C$-derivable formulas, and this could be
easily generalized to rules (in particular, modus ponens). However, we also need to establish these translations have
polynomial-size proofs. We cannot directly apply Lemma~\ref{lem:equi}: our translation does not commute with substitution,
thus instances of a fixed axiom do not translate to substitution instances of a single tautology. The claim below shows
the next best thing: they are instances of \emph{finitely many} tautologies, and this also applies to proper rules.
\begin{Cl}\label{cl:l1schema}
Let
\[\roo=\frac{\alpha_0(p_0,\dots,p_{k-1}),\dots,\alpha_{l-1}(p_0,\dots,p_{k-1})}{\beta(p_0,\dots,p_{k-1})}\]
be a Frege rule of~$\Fr{L^0_{C,\bot}}$. There is a set of $K\le 2^k$ rules $\{\roo_j:j<K\}$ derivable in~$L^1_C$
such that for any
substitution instance $\sigma(\roo)$ whose premises are $L^0$-tautologies, the rule $(\sigma(\roo))^*$ is a
substitution instance of some~$\roo_j$.
\end{Cl}
\begin{Pf*}
Let us consider substitution instances of formulas in variables $p_i$, $i<k$. It is clear from the definition that
if we fix the $k$~values $v(\sigma(p_i))\in\{0,1\}$, this determines $v(\sigma(\alpha))$ for every formula
$\alpha(p_0,\dots,p_{k-1})$. Moreover, in case this makes $v(\sigma(\alpha))=1$, the formula $(\sigma(\alpha))^*$ is
a substitution instance of a fixed formula: it can be shown by induction on the complexity of~$\alpha$
that $(\sigma(\alpha))^*=\sigma^*((\tau_I(\alpha))^*)$, where
$\tau_I(p_i)=\bot$ for $i\in I=\{i<k:v(\sigma(p_i))=0\}$, and $\sigma^*(p_i)=(\sigma(p_i))^*$ for each $i\notin I$.

Thus, let $K$ be the number of assignments to the variables $\{p_i:i<k\}$ that satisfy $\alpha_0,\dots,\alpha_{l-1}$, and $\{e_j:j<K\}$ be an
enumeration of such assignments. Put
$I_j=\{i<k:e_j(p_i)=0\}$, and
\[\roo_j=\frac{(\tau_{I_j}(\alpha_0))^*,\dots,(\tau_{I_j}(\alpha_{l-1}))^*}{(\tau_{I_j}(\beta))^*}.\]
Since $v(\tau_{I_j}(\alpha_i))=e_j(\alpha_i)=1$, and
\[\vdash_{L^1}\bigl(\tau_{I_j}(\alpha_0)\land\dots\land\tau_{I_j}(\alpha_{l-1})\to\tau_{I_j}(\beta)\bigr)^*\]
by Claim~\ref{cl:l1taut}, we have
\[\vdash_{L^1}(\tau_{I_j}(\alpha_0))^*\land\dots\land(\tau_{I_j}(\alpha_{l-1}))\to(\tau_{I_j}(\beta))^*,\]
i.e., the rule $\roo_j$ is valid in~$L^1$. On the other hand, assume that $\sigma$ is a substitution
such that $L^0$ proves $\sigma(\alpha_0)$, \dots, $\sigma(\alpha_{l-1})$. Let $e_j$ be the assignment defined by
$e_j(p_i)=v(\sigma(p_i))$. Then the discussion above shows that $(\sigma(\roo))^*=\sigma^*(\roo_j)$.
\end{Pf*}
Resuming the proof of Theorem~\ref{thm:bot-top}, Claim~\ref{cl:l1schema} shows that all inference steps in~$\pi^*$ are
instances of a fixed finite set of $\Fr{L^1_C}$-derivable rules, and as such they can be implemented with linear-size
$\Fr{L^1_C}$-derivations using Lemma~\ref{lem:equi}, resulting in a polynomial-size $\Fr{L^1_C}$ proof of the original formula~$\fii_m$.

This
completes the proof of Theorem~\ref{thm:bot-top} for Frege systems.
The argument for $\M{EF}$ is similar. If we formulate it as circuit Frege, we can use literally the same proof
as above. In the original formulation with extension axioms, we define $v$ so that it satisfies all original variables
and all extension axioms; the definition of $\fii^*$ stays unchanged, so that if $q$ is an extension variable with
$v(q)=1$, we put $q^*=q$. In effect, an extension axiom $q\eq\alpha$ from~$\pi$ disappears in~$\pi^*$ if
$v(\alpha)=0$, and transforms into an extension axiom $q\eq\alpha^*$ if $v(\alpha)=1$. The rest of the argument works
the same as for Frege.
\end{Pf}

\begin{Exm}\label{exm:bot-transl}
The conditions of Theorem~\ref{thm:bot-top} hold for the pair of logics $L^0=\lgc{KC}$, $L^1=\IPC$, or for
$L^0=L^1=\lgc{LC}$. 
\end{Exm}

\subsection{Elimination of $\land$}\label{sec:elim-land}

In this section, we explore to what extent we can eliminate conjunctions from proofs by means of the following simple
idea: by Lemma~\ref{lem:conj-free}, we can rewrite any formula in a proof as a conjunction of $\land$-free formulas, and
then we can replace it by the sequence of its conjuncts.
There are two principal difficulties with this method:
\begin{enumerate}
\item The construction in Lemma~\ref{lem:conj-free} substantially changes the shape and structure of the formula. In
particular, it does not commute with substitution: if we fix an axiom~$\Phi$, then $\land$-free translations of
substitution instances of~$\Phi$ are not themselves substitution instances of a single formula (or a small finite set).
This is responsible for the phenomenon in Proposition~\ref{prop:frag}, but even if we make sure that $\Phi$ does properly
axiomatize the relevant $\land$-free fragment, it means that there is no a priori reason why translation of instances
of~$\Phi$ should have poly-size proofs, and we will invest a lot of effort to fix this. It also means that the
construction will not work at all for $\M{SF}$ systems.
\item\label{item:18}
On the face of it, Lemma~\ref{lem:conj-free} exponentially blows up the sizes of formulas, hence it is no good for
polynomial simulations.
\end{enumerate}
Concerning~\ref{item:18}: first, Proposition~\ref{thm:moncon} exhibits a simple $\{\land,\lor\}$-formula that requires
exponentially many distinct conjuncts when written as a conjunction of $\land$-free formulas. Thus, the method does not
stand a chance in the presence of~$\lor$, and we will only work with $\lor$-free proofs. (In fact, we will mostly work
in the $\{\to,\land\}$-fragment, and indicate at the end how to extend the simulation to the
$\{\to,\land,\bot\}$-fragment.) In particular, our result will only apply to cofinal subframe (i.e.,
$\{\to,\bot\}$-axiomatized) logics.

Second, Corollary~\ref{cor:imp-conj} shows that an $\{\to,\land\}$-formula may still require an exponential blow-up when
written as a \emph{formula} with conjunctions only at top. Thus, even in this restricted case, our only hope for a
polynomial simulation is to work with \emph{circuits}. In particular, the method will only apply to $\M{EF}$, not
Frege. These constraints put us on the right track:
\begin{Lem}\label{lem:imp-conj-elim}
Given an $\{\to,\land\}$-circuit $C$ in variables $\{p_i:i<n\}$, we can construct in polynomial time a sequence of
$\{\to\}$-circuits $\{C^{p_i}:i<n\}$ such that
\[\vdash_\IPC C\eq\ET_{i<n}C^{p_i},\]
and each $C^{p_i}$ has the form $\Gamma\to p_i$, i.e., $H(C^{p_i})=p_i$.
\end{Lem}
\begin{Pf}
We construct the circuits $D^{p_i}$ for each $i<n$ and each circuit $D$ represented by a node of~$C$ by the following
induction on depth.

If $D=p_j$ is a variable, we may put $D^{p_j}=p_j$, and $D^{p_i}=p_i\to p_i$ for $i\ne j$.

If $D=D_0\to D_1$, we can use the equivalence
\[\vdash_\IPC D\eq\Bigl(\ET_{i<n}D_0^{p_i}\to\ET_{i<n}D_1^{p_i}\Bigr)\eq\ET_{i<n}(D_0^{p_0}\to\dots\to D_0^{p_{n-1}}\to D_1^{p_i}).\]
If $D=D_0\land D_1$, we have
\[\vdash_\IPC D\eq\Bigl(\ET_{i<n}D_0^{p_i}\land\ET_{i<n}D_1^{p_i}\Bigr)\eq\ET_{i<n}(D_0^{p_i}\land D_1^{p_i}).\]
Since $D_0^{p_i}$ and $D_1^{p_i}$ are of the form $\Gamma\to p_i$, this means
\[\vdash_\IPC D\eq\ET_{i<n}\bigl((D_0^{p_i}\to D_1^{p_i}\to p_i)\to p_i\bigr)\]
using~\eqref{eq:19}.

It is readily seen that we added at most $O(n^2)$ new gates for each node of~$C$, hence the total size of $\ET_iC^{p_i}$ is
$O(n^2\lh C)$, and it can be computed by a polynomial-time algorithm.
\end{Pf}

\begin{Lem}\label{lem:conj-elim-impl}
If $C$ is an implicational circuit with head $H(C)=p_h$, there are polynomial-time constructible
$\CF{\IPC_\to}$ proofs of the circuits
\[C^{p_h}\to C,\qquad C\to C^{p_h},\qquad C^{p_i}\]
for $i\ne h$.
\end{Lem}
\begin{Pf}
By straightforward induction on the construction of the circuit, using Lemma~\ref{lem:struct}. For the induction step, let
$C=C_0\to C_1$, in which case
\[C^{p_i}=C_0^{p_0}\to\dots\to C_0^{p_{n-1}}\to C_1^{p_i}.\]
By the induction hypothesis, we have constructed proofs of all but one of the circuits $C_0^{p_j}$, and of the
equivalence of the remaining one with~$C_0$. Using Lemma~\ref{lem:struct}, we can thus construct proofs of
\[C^{p_i}\ieq(C_0\to C_1^{p_i}).\]
For $i\ne h$, we also have a proof of~$C_1^{p_i}$, from which we derive $C^{p_i}$. For $i=h$, we have proofs of
$C_1^{p_h}\ieq C_1$, using which we derive $C^{p_h}\ieq(C_0\to C_1)$.
\end{Pf}

On the level of proofs, we first show that the translation works in the base case of intuitionistic logic, where we
need not worry about translations of unforeseen axioms.
\begin{Prop}\label{prop:conj-elim-weak}
Given an $\EF\IPC$ proof of an implicational formula~$\fii$, we can construct in polynomial time an
$\EF{\IPC_\to}$ proof of~$\fii$.
\end{Prop}
\begin{Pf}
It will be more convenient to work with the equivalent standard $\CF\IPC$ system. 
By Theorem~\ref{prop:lor-bot-pfs}, we may assume the proof only employs the connectives $\to,\land$.
Thus, let $C_0,\dots,C_m$ be an
$\CF{\IPC_{\to,\land}}$ proof of~$\fii=C_m$ in the axiom system given by \eqref{eq:41}--\eqref{eq:30}. Using the
transformation from Lemma~\ref{lem:imp-conj-elim}, we construct the sequence of implicational circuits
\[C_0^{p_0},\dots,C_0^{p_{n-1}},C_1^{p_0},\dots,C_m^{p_0},\dots,C_m^{p_{n-1}},\]
and we complete it to an $\CF{\IPC_\to}$ proof as follows.

If $C_i$ is inferred from $C_k=C_l\to C_i$ and~$C_l$ by modus ponens, $C_k^{p_j}$ is
\[C_l^{p_0}\to\dots\to C_l^{p_{n-1}}\to C_i^{p_j}.\]
Thus, we can infer $C_i^{p_j}$ from $C_k^{p_j}$ and $C_l^{p_0},\dots,C_l^{p_{n-1}}$ by $n$~applications of modus ponens.

If $C_i$ is an instance of axiom~\eqref{eq:18}, $C_i^{p_j}$ is of the form $\Gamma\to\gamma^{p_j}$, where $\Gamma$ is
the set of formulas
\[\bigl\{\alpha^{p_k},\alpha^{p_0}\to\dots\to\alpha^{p_{n-1}}\to\beta^{p_k},
     \alpha^{p_0}\to\dots\to\alpha^{p_{n-1}}\to\beta^{p_0}\to\dots\to\beta^{p_{n-1}}\to\gamma^{p_k}:k<n\bigr\}\]
in an appropriate order. Using Lemma~\ref{lem:struct}, we can derive
\[\Gamma\to\beta^{p_k}\]
for each $k<n$, and then we derive $\Gamma\to\gamma^{p_j}$ as required.

If $C_i$ is an instance of axiom~\eqref{eq:32}, $C_i^{p_j}$ is of the form
\[\bigl((\alpha^{p_0}\to\beta^{p_0}\to p_0)\to p_0\bigr)\to\dots
  \to\bigl((\alpha^{p_{n-1}}\to\beta^{p_{n-1}}\to p_{n-1})\to p_{n-1}\bigr)\to\alpha^{p_j}.\]
Using Lemma~\ref{lem:struct}, this follows from
\[\bigl((\alpha^{p_j}\to\beta^{p_j}\to p_j)\to p_j\bigr)\to\alpha^{p_j},\]
which is an instance of the implicational tautology
\[\bigl(((x\to z)\to y\to z)\to z\bigr)\to x\to z\]
as $\alpha^{p_j}$ is of the form $\cdots\to p_j$. Axioms \eqref{eq:31}, \eqref{eq:33}, and~\eqref{eq:30} are treated
similarly.

By Lemma~\ref{lem:conj-elim-impl}, we can complete the proof by inferring $\fii$ from the circuit~$C_m^{p_h}$.
\end{Pf}

Now we would like to generalize Proposition~\ref{prop:conj-elim-weak} to other si logics~$L$ axiomatized by implicational
formulas. Since our systems are necessarily finitely axiomatized, we may assume $L=\IPC+\Phi$, where $\Phi$ is an
implicational formula. In view of Proposition~\ref{prop:frag}, we need to recognize when $\IPC_\to+\Phi=(\IPC+\Phi)_\to$, as
otherwise we do not stand a chance of eliminating $\land$ from $(\IPC+\Phi)$-proofs at all, never mind the length.
\begin{Def}\label{def:conj-exp}
For any $n\ge1$, let $\xi_n$ denote the substitution
\[\xi_n(p_i)=\ET_{j<n}p_{in+j}.\]
Notice that up to associativity of~$\land$, $\xi_{nm}=\xi_n\circ\xi_m$.

Let $\Phi(p_0,\dots,p_{k-1})$ be an implicational formula with head $H(\Phi)=p_h$, and consider the transformation
from Lemma~\ref{lem:imp-conj-elim}. As in Lemma~\ref{lem:conj-elim-impl},
it is easy to see that the circuits $\bigl(\xi_n(\Phi)\bigr)^{p_{in+j}}$ have short $\CF{\IPC_\to}$ proofs for $i\ne
h$, hence there are short $\CF{\IPC_{\to,\land}}$ proofs of
\begin{equation}\label{eq:34}
\xi_n(\Phi)\eq\ET_{j<n}\bigl(\xi_n(\Phi)\bigr)^{p_{hn+j}}.
\end{equation}
Moreover, the circuits $\bigl(\xi_n(\Phi)\bigr)^{p_{hn+j}}$ for $j<n$ differ from each other only in inessential ways:
specifically, there are short $\CF{\IPC_\to}$ proofs of
\begin{equation}\label{eq:38}
\sigma_{j,j'}\bigl(\bigl(\xi_n(\Phi)\bigr)^{p_{hn+j}}\bigr)\ieq\bigl(\xi_n(\Phi)\bigr)^{p_{hn+j'}},
\end{equation}
where $\sigma_{j,j'}$ is the substitution swapping the variables $p_{hn+j}$ and~$p_{hn+j'}$.
In particular, proofs of $\bigl(\xi_n(\Phi)\bigr)^{p_{hn+j}}$ and $\bigl(\xi_n(\Phi)\bigr)^{p_{hn+j'}}$ are polynomial-time
constructible from each other. In light of this, we put
\[\Phi^{\land n}=\bigl(\xi_n(\Phi)\bigr)^{p_{hn}},\]
with the understanding that the choice of~$j=0$ here is arbitrary.
\end{Def}
\pagebreak[2]
\begin{Lem}\label{lem:impl-frag-char}
If $\Phi$ is an implicational formula, the following are equivalent.
\begin{enumerate}
\item\label{item:14}
$(\IPC+\Phi)_\to=\IPC_\to+\Phi$.
\item\label{item:15}
$(\IPC_{\to,\land}+\Phi)_\to=\IPC_\to+\Phi$.
\item\label{item:16}
$\IPC_\to+\Phi$ proves $\Phi^{\land n}$ for all $n\in\omega$.
\item\label{item:17}
$\IPC_\to+\Phi$ proves $\Phi^{\land2}$.
\end{enumerate}
\end{Lem}
\begin{Pf}

\ref{item:14}${}\eq{}$\ref{item:15}: We already observed that $(\IPC+\Phi)_{\to,\land}=\IPC_{\to,\land}+\Phi$ as a
consequence of Theorem~\ref{prop:lor-bot-pfs} (i.e., $\IPC$ is hereditarily $\{\to,\land\}$-conservative over
$\IPC_{\to,\land}$ in the terminology of Corollary~\ref{cor:her-cons}).

\ref{item:15}${}\eq{}$\ref{item:16}: The left-to-right implication follows from \eqref{eq:34}, as $\xi_n(\Phi)$ is
an instance of~$\Phi$ qua an axiom of $\IPC_{\to,\land}+\Phi$.

On the other hand, assume that an implicational formula~$\fii$ is provable in $\IPC_{\to,\land}+\Phi$. This means there
are substitutions $\sigma_u$, $u<r$, such that $\IPC_{\to,\land}$ proves
\[\ET_{u<r}\sigma_u(\Phi)\to\fii.\]
Each $\sigma_u(p_i)$ is equivalent to a conjunction of implicational formulas, hence we may assume
\[\sigma_u(p_i)=\ET_{j<n}\psi_{u,i,j}\]
with $\psi_{u,i,j}$ implicational, and $n$ the same for all $u$ and~$i$. But then
\[\sigma_u(\Phi)=\tau_u(\xi_n(\Phi)),\]
where $\tau_u(p_{in+j})=\psi_{u,i,j}$. Using~\eqref{eq:34}, the implicational formula
\[\{\tau_u\bigl(\bigl(\xi_n(\Phi)\bigr)^{p_{hn+j}}\bigr):u<r,j<n\}\to\fii\]
is provable in~$\IPC$, i.e., in~$\IPC_\to$. Using \eqref{eq:38}, each $\bigl(\xi_n(\Phi)\bigr)^{p_{hn+j}}$
is provable in $\IPC_\to+\Phi^{\land n}$, hence so is~$\fii$. By \ref{item:16}, it is also provable in $\IPC_\to+\Phi$.

\ref{item:16}${}\eq{}$\ref{item:17}: The left-to-right implication is trivial. Assume that $\IPC_\to+\Phi$ proves
$\Phi^{\land2}$, whence there are implicational substitutions $\sigma_u$, $u<r$, such that $\IPC$ proves
\[\ET_{u<r}\sigma_u(\Phi)\to\Phi^{\land2}.\]
By \eqref{eq:34} and~\eqref{eq:38}, $\IPC$ also proves
\begin{equation}\label{eq:35}
\ET_{u<r}\sigma_u(\Phi)\to\xi_2(\Phi).
\end{equation}
Let $n\ge1$, and apply $\xi_n$ to~\eqref{eq:35}. Since $\xi_n\circ\xi_2\equiv\xi_{2n}$, we obtain
\[\ET_{u<r}\xi_n\bigl(\sigma_u(\Phi)\bigr)\to\xi_{2n}(\Phi).\]
Applying \eqref{eq:34} to the implicational formulas $\sigma_u(p_i)$ in place of~$\Phi$, we see that each
$\xi_n(\sigma_u(p_i))$ is equivalent to a conjunction of $n$ implicational formulas. Thus,
\[\xi_n\circ\sigma_u\equiv\sigma_{n,u}\circ\xi_n\]
for some implicational substitution~$\sigma_{u,n}$, and $\IPC$ proves
\[\ET_{u<r}\sigma_{n,u}\bigl(\xi_n(\Phi)\bigr)\to\xi_{2n}(\Phi).\]

It follows by induction on~$n$ that there are implicational substitutions $\tau_{n,u}$, $u<r_n$, such that
$\IPC$ proves
\[\ET_{u<r_n}\tau_{n,u}(\Phi)\to\xi_n(\Phi),\]
hence also the implicational formula
\[\{\tau_{n,u}(\Phi):u<r_n\}\to\Phi^{\land n}.\]
Thus, $\IPC_\to+\Phi$ proves $\Phi^{\land n}$.
\end{Pf}

\begin{Exm}\label{exm:lc-impl}
G\"odel--Dummett logic $\lgc{LC}=\IPC+(p\to q)\lor(q\to p)$ can be axiomatized by the implicational formula
\[\Phi=\bigl((p\to q)\to r\bigr)\to\bigl((q\to p)\to r\bigr)\to r.\]
Up to renaming of variables, $\Phi^{\land2}$ is the formula
\begin{align*}
\bigl((p\to p'\to q)\to(p\to p'\to q')\to r\bigr)\to
\bigl((p\to p'\to q)\to(p\to p'\to q')\to r'\bigr)&\to\\
\bigl((q\to q'\to p)\to(q\to q'\to p')\to r\bigr)\to
\bigl((q\to q'\to p)\to(q\to q'\to p')\to r'\bigr)&\to r.
\end{align*}
We may simplify it using the substitution $r'\mapsto r$, as the result
\begin{equation}\label{eq:36}
\bigl((p\to p'\to q)\to(p\to p'\to q')\to r\bigr)\to
\bigl((q\to q'\to p)\to(q\to q'\to p')\to r\bigr)\to r
\end{equation}
still implies the original formula. We claim that \eqref{eq:36} is provable in $\IPC_\to+\Phi$, hence
$\lgc{LC}_\to=\IPC_\to+\Phi$ by Lemma~\ref{lem:impl-frag-char}. This can be shown by formalizing the argument below.

The effect of~$\Phi$ is that instead of proving a formula~$\fii$, it suffices to prove $(\alpha\to\beta)\to\fii$ and
$(\beta\to\alpha)\to\fii$ for a given pair of formulas $\alpha,\beta$; in other words, we may stipulate that
$\alpha\to\beta$ or $\beta\to\alpha$ is given as an assumption. By repeating this argument, we see that when proving
\eqref{eq:36} in $\IPC_\to+\Phi$, we may assume that a (fixed but arbitrary) linear order on the variables $p,p',q,q'$
is given. If the order is such that $\min\{p,p'\}\le\min\{q,q'\}$, then $p\to p'\to q$ and $p\to p'\to q'$ hold, hence the
first premise of~\eqref{eq:36} indeed implies~$r$. The other case is symmetric.
\qedhere\end{Exm}

The argument in the proof of Lemma~\ref{lem:impl-frag-char} shows that for any implicational axiom~$\Phi$, the
implicational fragment of the logic $\IPC+\Phi$ is axiomatized by
\[\IPC_\to+\{\Phi^{\land n}:n\in\omega\}.\]
However, this is an infinite set of axioms, hence it is no good as a basis for a Frege system.
\begin{Que}\label{que:fin-ax-impl}
Let $L$ be a si logic axiomatizable over~$\IPC$ by an implicational formula (i.e., a finitely axiomatizable subframe si
logic). Can it be axiomatized by a formula~$\Phi$ satisfying the conditions of Lemma~\ref{lem:impl-frag-char}? In other
words, is $L_\to$ finitely axiomatizable?
\end{Que}

\begin{Thm}\label{thm:conj-elim}
Let $L$ be a si logic axiomatizable by an implicational formula satisfying the conditions of
Lemma~\ref{lem:impl-frag-char} (in other words, a subframe si logic~$L$ such that $L_\to$ is finitely axiomatizable).

Given an $\EF L$ proof of an implicational formula~$\fii$, we can construct in
polynomial time an $\EF{L_\to}$ proof of~$\fii$.
\end{Thm}
\begin{Pf}
Write $L=\IPC+\Phi$, where the axiom $\Phi(p_0,\dots,p_{k-1})$ is as in Lemma~\ref{lem:impl-frag-char}.

We proceed in the same way as in the proof of Proposition~\ref{prop:conj-elim-weak}, using the same notation. The only
difference is that the proof being translated contains substitution instances of the new axiom, say
$C_i=\Phi(C_{i,0},\dots,C_{i,k-1})$. Now, in the bottom-up construction of $C_i^{p_j}$, we first construct
implicational circuits $C_{i,l}^{p_j}$ such that $C_{i,l}\equiv\ET_{j<n}C_{i,l}^{p_j}$, and these are then used as
black boxes in the construction of~$C_i^{p_j}$; thus, the construction proceeds as if $C_{i,l}$ were a conjunction of
$n$~new variables, which are substituted with $C_{i,l}^{p_j}$ in the end-result. The upshot is that
\[C_i^{p_j}=\sigma\bigl(\bigl(\xi_n(\Phi)\bigr)^{p_{hn+j}}\bigr),\]
where $\sigma(p_{ln+u})=C_{i,l}^{p_u}$, and $H(\Phi)=p_h$. Furthermore, $\bigl(\xi_n(\Phi)\bigr)^{p_{hn+j}}$ is
essentially a substitution instance of~$\Phi^{\land n}$ by~\eqref{eq:38}.

Thus, the only thing we need to do to complete the proof is to construct in polynomial time
$\CF{L_\to}$ proofs of~$\Phi^{\land n}$, given $n$ in unary. We will achieve this by analyzing the argument in
Lemma~\ref{lem:impl-frag-char} \ref{item:17}${}\to{}$\ref{item:16}.

As in~\eqref{eq:35}, let us fix $r\ge2$, implicational substitutions $\sigma_u$, $u<r$, and an
$\Fr{\IPC_{\to,\land}}$-derivation $\pi$ of $\xi_2(\Phi)$ from the set of assumptions $\{\sigma_u(\Phi):u<r\}$.

We may assume $n=2^t$ is a power of~$2$. For each
$s\le t$ and $u<r$, let $\sigma_{2^s,u}$ be the implicational substitution as in Lemma~\ref{lem:impl-frag-char}, so that
\begin{equation}\label{eq:37}
\xi_{2^s}\circ\sigma_u\equiv\sigma_{2^s,u}\circ\xi_{2^s}. 
\end{equation}
Since $\sigma_{2^s,u}(p_{i2^s+j})=\bigl(\xi_{2^s}(\sigma_u(p_i))\bigr)^{p_{h'2^s+j}}$ where $p_{h'}=H(\sigma_u(p_i))$, we
see readily that $\sigma_{2^s,u}$ is given by formulas of size $2^{O(s)}$, and there are $\Fr{\IPC_{\to,\land}}$ proofs
of~\eqref{eq:37} constructible in time $2^{O(s)}$.

By induction on $s\le t$, we construct $\CF{\IPC_{\to,\land}}$-derivations $\pi_s$ of $\xi_{2^s}(\Phi)$ from assumptions that are implicational
substitution instances of~$\Phi$ as follows.

Let $\pi_0$ be the trivial derivation of $\Phi$ from itself. Assume that $\pi_s$ has already been constructed, we will
build~$\pi_{s+1}$. We start by taking $\xi_{2^s}(\pi)$, which is a derivation of $\xi_{2^{s+1}}(\Phi)$ from assumptions
$\xi_{2^s}(\sigma_u(\Phi))$, $u<r$. Using short proofs of~\eqref{eq:37}, we include subderivations of every
$\xi_{2^s}(\sigma_u(\Phi))$ from $\sigma_{2^s,u}(\xi_{2^s}(\Phi))$. Finally, for each $u<r$, we include
$\sigma_{2^s,u}(\pi_s)$ as a derivation of $\sigma_{2^s,u}(\xi_{2^s}(\Phi))$.

Since we are working with circuits, we can write $\sigma_{2^s,u}(\pi_s)$ in such a way that it includes only one copy of
each substituted variable, hence roughly,
\[\lh{\sigma_{2^s,u}(\pi_s)}=\lh{\pi_s}+\sum_{\substack{i<k\\j<2^s}}\lh{\sigma_{2^s,u}(p_{i2^s+j})}.\]
Consequently,
\[\lh{\pi_{s+1}}=r\lh{\pi_s}+2^{O(s)},\]
and this recurrence resolves to
\[\lh{\pi_s}=2^{O(s)}.\]
For $s=t=\log_2n$, we obtain an $\CF{\IPC_{\to,\land}}$-derivation~$\pi_t$ of size $2^{O(t)}=n^{O(1)}$ of $\xi_n(\Phi)$
from implicational substitution instances of~$\Phi$, say $\{\tau_v(\Phi):v<m\}$. It is clear from the recursive
description that $\pi_t$ can also be constructed in polynomial time.

Using \eqref{eq:34} and the feasible deduction theorem, we can transform $\pi_t$ into an $\CF{\IPC_{\to,\land}}$ proof
of the implicational circuit
\[\{\tau_v(\Phi):v<m\}\to\Phi^{\land n}.\]
By Proposition~\ref{prop:conj-elim-weak}, we can transform it into an $\CF{\IPC_\to}$ proof, which we can turn into an
$\CF{(\IPC_\to+\Phi)}$ proof of $\Phi^{\land n}$ by modus ponens.
\end{Pf}

We can immediately generalize Theorem~\ref{thm:conj-elim} in the spirit of Theorem~\ref{prop:lor-bot-pfs}:
\begin{Cor}\label{cor:all-elim}
Let $P$ be a standard $\EF L$ system for a si logic~$L$ whose proper axiom is an implicational formula satisfying the
conditions of Lemma~\ref{lem:impl-frag-char}.

Given a $P$-proof $\pi$ of a formula~$\fii$, we can construct in polynomial time a $P$-proof $\pi'$ of~$\fii$ with
the following properties:
\begin{enumerate}
\item
The only conjunctions occurring in~$\pi'$ are subformulas of~$\fii$.
\item
The only disjunctions occurring in~$\pi'$ are subformulas of~$\fii$, and the only instances of~\eqref{eq:13} in~$\pi'$
are with $\gamma$ a subformula of~$\fii$ which is a variable, a disjunction, or~$\bot$.
\item
If $\fii$ is positive, then $\bot$ does not occur in~$\pi'$. Otherwise the only instances of~\eqref{eq:7} in~$\pi'$ are
with $\gamma$ a variable from~$\fii$.
\end{enumerate}
\end{Cor}
\begin{Pf}
Let $\hat\fii$ be the implicational formula from Corollary~\ref{cor:bot-lor-free}. We can construct in polynomial time its
$\EF L$ proof, hence also an $\EF{L_\to}$ proof by Theorem~\ref{thm:conj-elim}. We apply the substitution~$\sigma$ from
Corollary~\ref{cor:bot-lor-free} to this
proof. This reintroduces subformulas of~$\fii$, but no other formula fragments, hence the desired restrictions on
non-$\to$ connectives in the proof are satisfied. The proof now ends with a formula of the
form $\Delta\to\fii$, where $\Delta$ consists of instances of \eqref{eq:32}--\eqref{eq:7} satisfying the required
conditions; we include them as actual axioms, and derive~$\fii$ by modus ponens.
\end{Pf}

The whole construction can be easily adapted to $\{\to,\land,\bot\}$-circuits. In
Lemmas \ref{lem:imp-conj-elim} and~\ref{lem:conj-elim-impl} and Proposition~\ref{prop:conj-elim-weak} we handle $\bot$ as if it were an extra variable.
If $\Phi$ is a $\{\to,\bot\}$-formula whose head is a variable, Definition~\ref{def:conj-exp} works unchanged; if
$H(\Phi)=\bot$, we have
\[\xi_n(\Phi)\eq\bigl(\xi_n(\Phi)\bigr)^\bot\]
in place of~\eqref{eq:34}, and we duly define $\Phi^{\land n}=\bigl(\xi_n(\Phi)\bigr)^\bot$. (However, when this happens,
$\Phi$ is a negated formula, hence $\IPC+\Phi$ is either $\IPC$ or inconsistent; this is not an interesting case.)
The same argument as in Lemma~\ref{lem:impl-frag-char} yields:
\begin{Lem}\label{lem:impl-bot-frag-char}
If $\Phi$ is an $\{\to,\bot\}$-formula, the following are equivalent.
\begin{enumerate}
\item
$(\IPC+\Phi)_{\to,\bot}=\IPC_{\to,\bot}+\Phi$.
\item
$(\IPC_{\to,\land,\bot}+\Phi)_{\to,\bot}=\IPC_{\to,\bot}+\Phi$.
\item
$\IPC_{\to,\bot}+\Phi$ proves $\Phi^{\land n}$ for all $n\in\omega$.
\item
$\IPC_{\to,\bot}+\Phi$ proves $\Phi^{\land2}$.\noproof
\end{enumerate}
\end{Lem}
In the end, we obtain the following version of Corollary~\ref{cor:all-elim}.
\begin{Thm}\label{thm:all-but-bot}
Let $P$ be a standard $\EF L$ system for a si logic~$L$ whose proper axiom is an $\{\to,\bot\}$-formula satisfying the
conditions of Lemma~\ref{lem:impl-bot-frag-char}.

Given a $P$-proof $\pi$ of a formula~$\fii$, we can construct in polynomial
time a $P$-proof $\pi'$ of~$\fii$ with the following properties:
\begin{enumerate}
\item
The only conjunctions occurring in~$\pi'$ are subformulas of~$\fii$.
\item
The only disjunctions occurring in~$\pi'$ are subformulas of~$\fii$, and the only instances of~\eqref{eq:13} in~$\pi'$
are with $\gamma$ a subformula of~$\fii$ which is a variable, a disjunction, or~$\bot$.
\item
The only instances of~\eqref{eq:7} in~$\pi'$ are with $\gamma$ a variable from~$\fii$.\noproof
\end{enumerate}
\end{Thm}

\subsection{Exponential separation}\label{sec:expon-separ}

The exponential speed-up of $\M{SF}$ over~$\M{EF}$ exhibited in Corollary~\ref{cor:expsep} applies to~$\IPC$ and many other si
logics, but not to their \emph{fragments}, despite the tautologies being purely implicational. We will close this gap
by showing that the tautologies have short (poly-time constructible) $\SF{\IPC_\to}$ proofs. Notice that
Theorem~\ref{prop:lor-bot-pfs} (or rather its $\M{SF}$-version, as mentioned there) provides poly-time constructible
$\SF{\IPC_{\to,\land}}$ proofs of these tautologies. Unfortunately, elimination of~$\land$ in
Section~\ref{sec:elim-land} does not work for $\M{SF}$ systems, hence we will have to resort to an ad~hoc argument to
get it down all the way to~$\IPC_\to$. Similarly to the proofs of \cite[L.~6.26,29]{ej:sfef}, we will start from
classical proofs of the tautologies, and judiciously replace bits and pieces to obtain the desired result.

In order to extract $\IPC_\to$~proofs out of $\CPC$~proofs, we will employ an implicational version of the negative
(Glivenko) translation. Hopefully, this may be of independent interest, hence the argument is not \emph{completely}
ad~hoc after all.
\begin{Def}
The class of \emph{negative} $\{\to,\bot\}$-formulas is the smallest class of formulas such that
\begin{itemize}
\item if $\Gamma$ is a (possibly empty) sequence of variables and negative formulas, then $\Gamma\to\bot$ is a
negative formula.
\end{itemize}
Given a negative formula~$\fii$, and a variable~$u$, let $\fii^u$ denote the result of replacing $\bot$ with~$u$
in~$\fii$.
\end{Def}
\begin{Prop}\label{lem:neg}
Given a $\Fr\CPC$ proof of a negative formula~$\fii$, we can construct in polynomial time an $\Fr{\IPC_\to}$ proof of
$\fii^u$.
\end{Prop}
\begin{Pf}
It will be convenient to formulate classical logic with unbounded fan-in NAND as the only connective; we will write it
as $\ob\Phi$, where $\Phi$ is a sequence of formulas. It is easy to check that the one-sided sequent%
\footnote{Here, the meaning of a sequent $\Gamma=\p{\fii_0,\dots,\fii_{n-1}}$ is the \emph{conjunction}
$\ET_{i<n}\fii_i$. Think of it as representing the two-sided sequent $\Gamma\Longrightarrow{}$.}
calculus with the
following rules is a complete proof system for proving \emph{unsatisfiability} of sequences of NAND-formulas, p-equivalent to Frege:
\begin{itemize}
\item the usual structural rules of exchange, contraction, and weakening;
\item initial sequents $\ob\Phi,\Phi$;
\item the cut rule
\[\frac{\Gamma,\Phi\qquad\Gamma,\ob\Phi}\Gamma.\]
\end{itemize}
Now, if we identify $\ob\Phi$ with $\Phi\to\bot$, then NAND-formulas are exactly the variables and negative formulas.
If $\Gamma=\p{\fii_i:i<n}$, let us write $\Gamma^u=\p{\fii_i^u:i<n}$. Then given a proof
$\Gamma_0;\Gamma_1;\dots;\Gamma_n$ of the unsatisfiability of~$\Gamma=\Gamma_n$ in the above system, we consider the
sequence of formulas
\[\Gamma_0^u\to u;\dots;\Gamma_n^u\to u\]
and make it an $\Fr{\IPC_\to}$ proof of $\Gamma^u\to u$: structural rules are handled by
Lemma~\ref{lem:struct}~\ref{item:12}, initial sequents translate to the tautologies
\[(\Phi^u\to u)\to\Phi^u\to u,\]
and the cut rule translates to
\[\frac{\Gamma^u\to\Phi^u\to u\qquad\Gamma^u\to(\Phi^u\to u)\to u}{\Gamma^u\to u},\]
which has a short derivation by Lemma~\ref{lem:struct}~\ref{item:13}.

Finally, if $\fii$ is a negative formula, it has the form $\Gamma\to\bot$, where $\Gamma$ consists of negative formulas
and variables. A classical Frege proof of~$\fii$ can be transformed into an unsatisfiability proof for the
sequent~$\Gamma$, which we translate to an $\Fr{\IPC_\to}$ proof of the formula $(\Gamma^u\to u)=\fii^u$ as above.
\end{Pf}
\begin{Def}\label{def:monneg}
Given a monotone formula $\fii$, we define a negative formula~$\nneg\fii$ equivalent to $\neg\fii$ as follows:
\begin{align*}
\nneg p&=p\to\bot,\\
\nneg(\fii\lor\psi)&=(\nneg\fii\to\nneg\psi\to\bot)\to\bot,\\
\nneg(\fii\land\psi)&=(\nneg\fii\to\bot)\to(\nneg\psi\to\bot)\to\bot,\\
\nneg\bot&=\bot\to\bot,\\
\nneg\top&=\bot.
\end{align*}
\end{Def}

Using~\eqref{eq:19}, it is easy to see that $(\nneg\fii)^u$ is equivalent in~$\IPC$ to $\fii\to u$. Thus, it should be
obvious that the formulas below have short $\Fr\IPC$ proofs. However, this does not give $\Fr{\IPC_\to}$ proofs, as the
formulas $\fii\to u$ are meaningless as such in~$\IPC_\to$ (in fact, we essentially introduced $\nneg\fii$ and
$(\nneg\fii)^u$ to have a stand-in for $\fii\to u$). Thus, we need to work a little harder.
\begin{Lem}\label{lem:neg-mon-u}
Let $\fii$ and~$\psi$ be monotone formulas. There are polynomial-time constructible $\Fr{\IPC_\to}$ proofs of
\begin{gather}
\label{eq:25}\bigl(\bigl((\nneg\fii)^u\to u\bigr)\to u\bigr)\ieq(\nneg\fii)^u,\\
\label{eq:28}\bigl(\nneg(\fii\lor\psi)\bigr)^u\to(\nneg\fii)^u,\qquad\bigl(\nneg(\fii\lor\psi)\bigr)^u\to(\nneg\psi)^u,\\
\label{eq:29}(\nneg\fii)^u\to(\nneg\psi)^u\to\bigl(\nneg(\fii\lor\psi)\bigr)^u,\\
\label{eq:26}(\nneg\fii)^u\to\bigl(\nneg(\fii\land\psi)\bigr)^u,\qquad(\nneg\psi)^u\to\bigl(\nneg(\fii\land\psi)\bigr)^u,\\
\label{eq:24}(\nneg\fii)^{v\to u}\ieq\bigl(v\to(\nneg\fii)^u\bigr),\\
\label{eq:22}(v\to u)\to(\nneg\fii)^v\to(\nneg\fii)^u.
\end{gather}
\end{Lem}
\begin{Pf}
\eqref{eq:25} follows from~\eqref{eq:19}, as $(\nneg\fii)^u$ is of the form $\Gamma\to u$. Then it is straightforward
to show \eqref{eq:28}--\eqref{eq:26}.

\eqref{eq:24}: Clearly, $v\to\bigl((v\to u)\ieq u\bigr)$ gives short proofs of
\[v\to\bigl((\nneg\fii)^{v\to u}\ieq(\nneg\fii)^u\bigr).\]
Then we can derive
\[\bigl(v\to(\nneg\fii)^u\bigr)\ieq\bigl(v\to(\nneg\fii)^{v\to u}\bigr)\ieq(\nneg\fii)^{v\to u},\]
using the fact that $(\nneg\fii)^{v\to u}$ is of the form $\Gamma\to v\to u$.

\eqref{eq:22}: For any occurrence of a subformula $\psi\sset\fii$, we define a set of formulas
$\Gamma_\psi$ by top-to-bottom induction as follows. We put $\Gamma_\fii=\nul$. If $\psi=\psi_L\lor\psi_R$, then
$\Gamma_{\psi_L}=\Gamma_{\psi_R}=\Gamma_\psi$. If $\psi=\psi_L\land\psi_R$, we put
$\Gamma_{\psi_L}=\Gamma_\psi$, and $\Gamma_{\psi_R}=\Gamma_\psi\cup\{\psi_L\}$. Let
\begin{align*}
\Gamma_\psi^u&=\bigl\{(\nneg\chi)^u\to u:\chi\in\Gamma_\psi\bigr\},\\
\psi_*^u&=\Gamma_\psi^u\to(\nneg\psi)^u.
\end{align*}
We will construct proofs of
\begin{equation}\label{eq:27}
(v\to u)\to\psi_*^v\to\psi_*^u
\end{equation}
by induction on~$\psi\sset\fii$, where we order the subformulas so that the left child of~$\psi$ and its subformulas
come before the right child and its subformulas, which come before~$\psi$. In particular, all formulas in~$\Gamma_\psi$
come before~$\psi$. Notice that for $\psi=\fii$, \eqref{eq:27} is just~\eqref{eq:22}.

Let $\psi=p$ be a variable. If $\Gamma_\psi$ is empty, \eqref{eq:27} is
\[(v\to u)\to(p\to v)\to(p\to u).\]
Otherwise $\Gamma_\psi=\{\chi\}\cup\Gamma_\chi$ for some~$\chi$ coming before~$\psi$. We have
\[\psi_*^u\ieq\bigl(\Gamma_\chi^u\to\bigl((\nneg\chi)^u\to u\bigr)\to p\to u\bigr)
\ieq\bigl(p\to\Gamma_\chi^u\to(\nneg\chi)^u\bigr)\ieq(p\to\chi_*^u)\]
by~\eqref{eq:25}, hence \eqref{eq:27} follows from the induction hypothesis for~$\chi$.

If $\psi=\psi_L\land\psi_R$, then
\[(\psi_R)_*^u\ieq\bigl(\Gamma_\psi^u\to\bigl((\nneg\psi_L)^u\to u\bigr)\to(\nneg\psi_R)^u\bigr)
\ieq\bigl(\Gamma_\psi^u\to(\nneg\psi)^u\bigr)\ieq\psi_*^u,\]
hence the induction hypothesis for $\psi_R$ immediately yields \eqref{eq:27} for~$\psi$.

If $\psi=\psi_L\lor\psi_R$, we have
\[(\psi_L)_*^u\ieq\bigl(\Gamma_\psi^u\to(\nneg\psi_L)^u\bigr),\qquad
(\psi_R)_*^u\ieq\bigl(\Gamma_\psi^u\to(\nneg\psi_R)^u\bigr).\]
Thus, we can derive
\begin{gather*}
(v\to u)\to\bigl(\Gamma_\psi^v\to(\nneg\psi)^v\bigr)\to\bigl(\Gamma_\psi^u\to(\nneg\psi_L)^u\bigr)\\
(v\to u)\to\bigl(\Gamma_\psi^v\to(\nneg\psi)^v\bigr)\to\bigl(\Gamma_\psi^u\to(\nneg\psi_R)^u\bigr)
\end{gather*}
using \eqref{eq:28} and the induction hypothesis, whence
\[(v\to u)\to\bigl(\Gamma_\psi^v\to(\nneg\psi)^v\bigr)\to\bigl(\Gamma_\psi^u\to(\nneg\psi)^u\bigr)\]
using~\eqref{eq:29}.
\end{Pf}

\begin{Thm}\label{thm:sep-imp}
There is a sequence of implicational formulas that have polynomial-time constructible
$\SF{\IPC_\to}$ proofs, but require $\EF L$ proofs of size $2^{n^{\Omega(1)}}$ for any si logic $L$ with unbounded
branching.
\end{Thm}
\begin{Pf}
A somewhat simplified version of the implicational tautologies $\fii_n$ from Corollary~\ref{cor:expsep}, translating the
tautologies~\eqref{eq:2} from Theorem~\ref{thm:sfef-lb}, can be written as
\begin{multline}\label{eq:6}
\Bigl(\Bigl(\nneg\ET_{i<n}(p_i\lor p'_i)\Bigr)^u\to u\Bigr)\to
\Bigl(\Bigl(\nneg\ET_{i<n}(s_i\lor s'_i)\Bigr)^v\to u\Bigr)\to
\Bigl(\Bigl(\nneg\ET_{i<n}(r_i\lor r'_i)\Bigr)^w\to u\Bigr)\to\\
\bigl(\nneg\gamma(\vec p,\vec s,\vec{s'})\bigr)^v\to
\bigl(\nneg\delta(\vec{p'},\vec r,\vec{r'})\bigr)^w\to
u,
\end{multline}
where $\gamma$ and~$\delta$ are monotone. Since \eqref{eq:2} has short classical Frege proofs, so does
\[\neg\nneg\ET_{i<n}(p_i\lor p'_i)\to\neg\nneg\ET_{i<n}(s_i\lor s'_i)\to\neg\nneg\ET_{i<n}(r_i\lor r'_i)\to
  \nneg\gamma(\vec p,\vec s,\vec{s'})\to\nneg\delta(\vec{p'},\vec r,\vec{r'})\to\bot,\]
hence using Proposition~\ref{lem:neg}, we can construct short $\Fr{\IPC_\to}$ proofs of
\begin{multline}\label{eq:20}
\Bigl(\Bigl(\nneg\ET_{i<n}(p_i\lor p'_i)\Bigr)^u\to u\Bigr)\to
\Bigl(\Bigl(\nneg\ET_{i<n}(s_i\lor s'_i)\Bigr)^u\to u\Bigr)\to
\Bigl(\Bigl(\nneg\ET_{i<n}(r_i\lor r'_i)\Bigr)^u\to u\Bigr)\to\\
\bigl(\nneg\gamma(\vec p,\vec s,\vec{s'})\bigr)^u\to
\bigl(\nneg\delta(\vec{p'},\vec r,\vec{r'})\bigr)^u\to
u.
\end{multline}
Putting
\[\Gamma(\vec p,\vec{p'},\vec r,\vec{r'},u)=\Bigl\{\Bigl(\nneg\ET_{i<n}(p_i\lor p'_i)\Bigr)^u\to u,
\Bigl(\nneg\ET_{i<n}(r_i\lor r'_i)\Bigr)^u\to u,
(\nneg\delta)^u\Bigr\},\]
we derive
\[\Gamma\to(\nneg\gamma)^v\to\Bigl(\Bigl(\nneg\ET_{i<n}(s_i\lor s'_i)\Bigr)^u\to u\Bigr)\to(v\to u)\to u\]
using~\eqref{eq:22}. We proceed to prove
\begin{equation}\label{eq:23}
\Gamma\to(\nneg\gamma)^v\to\Bigl(\Bigl(\nneg\ET_{j\le i<n}(s_i\lor s'_i)\Bigr)^u\to u\Bigr)
\to\Bigl(\Bigl(\nneg\ET_{i<j}(s_i\lor s'_i)\Bigr)^v\to u\Bigr)\to u
\end{equation}
by induction on~$j\le n$. Assume we have~\eqref{eq:23} for~$j$. Note that
$\bigl(\nneg\ET_{j\le i<n}(s_i\lor s'_i)\bigr)^u$ is equivalent to
\[\bigl((s_j\to u)\to(s'_j\to u)\to u\bigr)\to\Bigl(\nneg\ET_{j<i<n}(s_i\lor s'_i)\Bigr)^u.\]
Thus, if we substitute $\top$ for~$s_j$, and $s_j\to v$ for~$v$, we obtain
\[\Gamma\to\bigl(\nneg\gamma(s_j/\top)\bigr)^{s_j\to v}\to\Bigl(\Bigl(\nneg\ET_{j<i<n}(s_i\lor s'_i)\Bigr)^u\to u\Bigr)
\to\Bigl(\Bigl(\nneg\ET_{i<j}(s_i\lor s'_i)\Bigr)^{s_j\to v}\to u\Bigr)\to u.\]
(It is essential here that neither $s_j$ nor~$v$ occur in~$\Gamma$.)
Using~\eqref{eq:24} and short proofs of
\[(\nneg\gamma)^v\to s_j\to\bigl(\nneg\gamma(s_j/\top)\bigr)^v,\]
we obtain
\[\Gamma\to(\nneg\gamma)^v\to\Bigl(\Bigl(\nneg\ET_{j<i<n}(s_i\lor s'_i)\Bigr)^u\to u\Bigr)
\to\Bigl[\Bigl(s_j\to\Bigl(\nneg\ET_{i<j}(s_i\lor s'_i)\Bigr)^v\Bigr)\to u\Bigr]\to u.\]
A symmetric argument gives the same formula with $s'_j$ in place of~$s_j$. Putting $\fii=\ET_{i<j}(s_i\lor s'_i)$, we
have
\[\bigl(s_j\to(\nneg\fii)^v\bigr)\to\bigl(s'_j\to(\nneg\fii)^v\bigr)\to
  \underbrace{\bigl((\nneg\fii)^v\to v\bigr)\to\bigl((s_j\to v)\to(s'_j\to v)\to v\bigr)\to v}
   _{\textstyle\bigl(\nneg\ET_{i\le j}(s_i\lor s'_i)\bigr)^v},\]
and using
\[x\to y\to z\vdash((x\to u)\to u)\to((y\to u)\to u)\to(z\to u)\to u,\]
we can derive
\[\Gamma\to(\nneg\gamma)^v\to\Bigl(\Bigl(\nneg\ET_{j<i<n}(s_i\lor s'_i)\Bigr)^u\to u\Bigr)
\to\Bigl(\Bigl(\nneg\ET_{i\le j}(s_i\lor s'_i)\Bigr)^v\to u\Bigr)\to u,\]
which is \eqref{eq:23} for $j+1$.

In the end, $j=n$ gives
\[\Gamma\to(\nneg\gamma)^v\to\Bigl(\Bigl(\nneg\ET_{i<n}(s_i\lor s'_i)\Bigr)^v\to u\Bigr)\to u.\]
Repeating the same argument with $\delta,w,r_i,r'_i$ in place of $\gamma,v,s_i,s'_i$ yields~\eqref{eq:6}.
\end{Pf}

\section{Conclusion}\label{sec:conclusion}
The diverse results in this paper show that by and large, the proof complexity of common calculi for intuitionistic
logic is concentrated on implicational formulas. On the one hand, we do not miss anything substantial by restricting
attention to the lengths of proofs of purely implicational tautologies. On the other hand, we may assume intuitionistic
proofs to come in a sort of normal form where the bulk of the proof takes place in the implicational fragment
of~$\IPC$, except for a small predetermined set of nonimplicational axioms that essentially reconstruct the
nonimplicational subformulas of the tautology being proved. To some extent the same is true of other
superintuitionistic logics.

Nevertheless, our work inevitably revealed various rough spots where our methods leave something to be desired,
especially in connection with extensions of intuitionistic logic. In particular, our results on elimination of
conjunctions from proofs in Section~\ref{sec:elim-land} may be limited more by the specific method we employ rather
than intrinsically:
\begin{Que}\label{que:conj-noncsf}
Can we extend Theorem~\ref{thm:conj-elim} to si logics whose axioms involve disjunctions, under suitable conditions on
their axioms?
\end{Que}
\begin{Que}\label{que:sf-conj}
Can we extend Theorem~\ref{thm:conj-elim} to $\SF\IPC$ or other substitution Frege systems?
\end{Que}
Recall that we stated another problem as Question~\ref{que:fin-ax-impl}.

For Question~\ref{que:sf-conj}, it is unclear whether we should expect a positive or a negative answer. Indeed, a
superpolynomial speedup of $\SF\IPC$ over $\SF{\IPC_\to}$ would be an exciting result. Likewise, it would be very
interesting if we could turn some of the lower bounds in Appendix~\ref{sec:cxt-flas} into lower bounds on the lengths
of proofs.

Let us also mention the following open problems in the area of intuitionistic proof complexity.
\begin{Que}
Does $\Fr\IPC$ polynomially simulate $\EF\IPC$?
\end{Que}
\begin{Que}
Can we prove an unconditional lower bound on $\SF\IPC$ proofs?
\end{Que}
While neither question is particularly connected to implicational fragments, we note that they would be interesting
to resolve even for proper fragments of~$\IPC$, or another si logic.

\subsection*{Acknowledgements}
I am indebted to Pavel Hrube\v s for suggesting the topic of this paper, and fruitful discussion, in particular drawing
my attention to the constructions in Sections \ref{sec:elim-bot} and~\ref{sec:elim-land}. I also thank the anonymous
referee for useful suggestions to improve the paper.

The research leading to these results has received funding from the European Research Council under the European
Union's Seventh Framework Programme (FP7/2007--2013)~/ ERC grant agreement no.~339691. The Institute of Mathematics of
the Czech Academy of Sciences is supported by RVO: 67985840.

\providecommand{\bysame}{\leavevmode\hbox to5em{\hrulefill}\thinspace}

\appendix
\section{Axiomatization of fragments}\label{sec:axiom-fragm}

This section makes an exposition of a result on fragments of si logics originally due to
Wro\'nski~\cite{wron:red}, related to elimination of connectives from Section~\ref{sec:elim-proof}.

We are interested in the following question: if $C$ is a set of connectives, is it true that whenever $L$ is a si logic
axiomatized over~$\IPC$ by a $C$-formula $\Phi$, the $C$-fragment of $L$ is axiomatized by $\Phi$ over $\IPC_C$? 
In other words, is every axiomatic extension of~$\IPC_C$ the $C$-fragment of a si logic? While
trivially $\IPC_C+\Phi\sset(\IPC+\Phi)_C$, this inclusion may a~priori be strict: a proof of a $C$-formula $\fii$ in
$\IPC+\Phi$ may use substitution instances $\sigma(\Phi)$ involving arbitrary connectives, whereas a proof in
$\IPC_C+\Phi$ only allows substitutions by $C$-formulas. Let us first properly introduce some terminology.
\begin{Def}\label{def:her-cons}
Let $C,C_0,C_1\sset C_\IPC$ be such that ${\to}\in C\sset C_0\cap C_1$. We say that \emph{$\IPC_{C_0}$ is hereditarily
$C$-conservative over $\IPC_{C_1}$} if for all sets of $C$-formulas $X$, the $C$-fragment of $\IPC_{C_0}+X$ is included
in $\IPC_{C_1}+X$.
\end{Def}
\begin{Obs}\label{obs:incl}
Let $C,C_0,C_1$ be as in Definition~\ref{def:her-cons}. If $C_0\sset C_1$, then $\IPC_{C_0}$ is hereditarily $C$-conservative
over $\IPC_{C_1}$.

Let $C',C'_0,C'_1$ also obey the restrictions from Definition~\ref{def:her-cons}, and $C'\sset C$, $C'_0\sset C_0$,
and $C_1\sset C'_1$. If $\IPC_{C_0}$ is hereditarily $C$-conservative over $\IPC_{C_1}$, then $\IPC_{C'_0}$ is
hereditarily $C'$-conservative over $\IPC_{C'_1}$.
\noproof\end{Obs}

A partial positive answer is provided by Theorem~\ref{prop:lor-bot-pfs}, which shows that we do not need substitutions with
$\lor$ or~$\bot$ if these connectives do not appear in $\Phi$ or~$\fii$:
\begin{Cor}\label{cor:her-cons}
If $C,C_0,C_1$ are as in Definition~\ref{def:her-cons}, and ${\land}\in C_1$, then $\IPC_{C_0}$ is hereditarily
$C$-conservative over $\IPC_{C_1}$.
\noproof\end{Cor}

However, these results for $\lor$ and~$\bot$ cannot be directly generalized to~$\land$---in fact,
Observation~\ref{obs:incl} and Corollary~\ref{cor:her-cons} cannot be improved:
\begin{Thm}[Wro\'nski \cite{wron:red}]\label{thm:frag-all}
Let $C,C_0,C_1$ be as  in Definition~\ref{def:her-cons}. The following are equivalent:
\begin{enumerate}
\item $\IPC_{C_0}$ is hereditarily $C$-conservative over $\IPC_{C_1}$.
\item ${\land}\in C_1$ or $C_0\sset C_1$.
\end{enumerate}
\end{Thm}
(Wro\'nski only explicitly mentions the cases with $C=C_1$, but his argument proves the formulation given here. He also
considers fragments without implication, which we are not interested in.)

For completeness, we will prove Theorem~\ref{thm:frag-all} below. In view of Observation~\ref{obs:incl} and Corollary~\ref{cor:her-cons}, it
suffices to show the three non-conservativity results in the next proposition.%
\footnote{We mention that item~\ref{item:10} is relevant to \cite{cin-met:intfrag}.}
We assume familiarity with basic universal algebra and algebraic methods in propositional logics, see e.g.\
\cite{bur-san,blok-pig}. For typographic convenience, we will write $L\vdash\fii$ instead of $\vdash_L\fii$.
\begin{Prop}\label{prop:frag}
\ \begin{enumerate}
\item\label{item:9}
$\IPC_{\to,\land}$ is not hereditarily $\to$-conservative over~$\IPC_{\to,\lor,\bot}$.
\item\label{item:11}
$\IPC_{\to,\bot}$ is not hereditarily $\to$-conservative over~$\IPC_{\to,\lor}$.
\item\label{item:10}
$\IPC_{\to,\lor}$ is not hereditarily $\to$-conservative over~$\IPC_{\to,\bot}$.
\end{enumerate}
\end{Prop}
\begin{Pf}
We will work for the moment with the implicational fragment.

$\IPC_\to$ is regularly algebraizable wrt the class of $\to$-subreducts of Heyting algebras, which is the variety of
$\bckw$ (aka Hilbert) algebras, hence axiomatic extensions of~$\IPC_\to$ correspond to subvarieties $V\sset\bckw$. By
Diego's theorem \cite[\S5.4]{cha-zax}, $\bckw$ is locally finite, hence $V$ is generated by its finite subdirectly irreducible (sdi)
algebras. Congruences on a $\bckw$-algebra $\p{A,1,\to}$ are determined by filters (subsets $F\sset A$ such that $1\in
F$, and $a,a\to b\in F$ implies $b\in F$). Every element $a\in A$ generates the principal filter $a\up$. It follows
that $A$ is sdi iff it has an \emph{opremum}: a largest element strictly below~$1$.

Let $A$ be a finite sdi $\bckw$-algebra with opremum~$o$. The \emph{characteristic formula of~$A$} is the formula
\[\fii_A=\Xi_A\to p_o\]
in variables $\{p_a:a\in A\}$, where the set of formulas~$\Xi_A$ consists of
\[(p_a\to p_b)\to p_{a\to b},\qquad p_{a\to b}\to(p_a\to p_b)\]
for $a,b\in A$. Clearly, $\fii_A$ is invalid in~$A$ under the valuation $v(p_a)=a$. On the other hand, let $v$ be a
valuation in a $\bckw$-algebra $B$ such that $v(\fii_A)\ne1$. The filter generated by~$v(\Xi_A)$, which is
\[F=\{b\in B:v(\Xi_A)\to b=1\},\]
does not contain~$v(p_o)$, hence $v$ induces a valuation $v_F$ on the quotient algebra $B/F$ such that $v_F(\Xi_A)=1$,
and $v_F(p_o)<1$. Then $v_F(p_{a\to b})=v_F(p_a)\to v_F(p_b)$ by the definition of~$\Xi_A$, hence the mapping $f\colon
A\to B/F$ defined by $f(a)=v_F(p_a)$ is a homomorphism. Its kernel does not contain the generator~$o$ of the least
nontrivial congruence on~$A$, hence $f$ is in fact an embedding. Thus, we have established that $B\nmodel\fii_A$ iff
$A\in\clsop S(\clsop H(B))$.

With $A$ as above, let $B$ be a sdi algebra in $\clsop H(\clsop S(\clsop P(A)))$. If $B$ is finite, $A\nmodel\fii_B$,
hence $B\in\clsop S(\clsop H(A))$.
If $B$ were infinite, let $B'$ be its finite(ly generated) subalgebra including the opremum.
Then $B'\in\clsop S(\clsop H(A))$ by the previous argument, which gives a contradiction if we choose $\lh{B'}>\lh A$.%
\footnote{Alternatively, we could show that $\bckw$ is a congruence-distributive variety, and apply J\'onsson's lemma
for the slightly weaker conclusion $B\in\clsop H(\clsop S(A))$.}

Now, assume that $A$ is a finite sdi $\IPC_{C_1}$-algebra, and that $\IPC_{C_0}$ is hereditarily
$\to$-conservative over~$\IPC_{C_1}$. Let $X$ be the set of $\to$-formulas valid in~$A$. We have
$\IPC_{C_1}+X\nvdash\fii_A$ as witnessed by~$A$, hence $\IPC_{C_0}+X\nvdash\fii_A$, which means that there exists an
$\IPC_{C_0}$-algebra $B$ such that $B\res{\to}\in H(S(P(A\res{\to})))$, and $B\nmodel\fii_A$. We may assume $B$ is sdi.
Since $B\res{\to}$ and $B$ have the same congruences, $B\res{\to}$ is also sdi.
By the above, $B\res{\to}\in S(H(A\res{\to}))$; on the other hand,
$B\nmodel\fii_A$ implies $A\res{\to}\in S(H(B\res{\to}))$, and as $A$ is finite, this gives
$A\res{\to}\simeq B\res{\to}$. In other words, $A$ itself must be an $\IPC_{C_0}$-algebra to begin
with. (Note that $\to$ determines~$\le$, hence also the lattice operations $\land,\lor,\bot$. Thus, $A\res{\to}$ has at
most one expansion to an $\IPC_{C_0}$-algebra.)

It remains to find specific counterexamples for the three statements.

\ref{item:9}: Let $B$ be the Heyting algebra with domain $\{1,o,a,b,c,d,\bot\}$, where $\bot<c<b<o<1$,
$\bot<d<a<o$, and $d<b$. Put $A=B\bez\{d\}$, and note that $A$ is a $\{\to,\lor,\bot\}$-subalgebra of~$B$,
and it is sdi with opremum~$o$. However, $A$ is not an $\IPC_{\to,\land}$-algebra, being a non-distributive
lattice.

For \ref{item:11}, let $B$ be the Heyting algebra with domain $\{1,o,a,b,\bot\}$, where $\bot<\{a,b\}<o<1$, and $A$
be its $\{\to,\lor\}$-subalgebra $B\bez\{\bot\}$, which is sdi, and has no least element.

For \ref{item:10}, we can use $B=\{1,o,a,b,m,c,d,\bot\}$, where $\bot<\{c,d\}<m<\{a,b\}<o<1$, and $A=B\bez\{m\}$, which
is a $\{\to,\bot\}$-subalgebra of~$B$, but not an $\IPC_{\to,\lor}$-algebra, as $c$ and~$d$ have no join in~$A$.
\end{Pf}
\begin{Exm}\label{exm:incomp-frag}
To illustrate the abstract nonsense with actual formulas, put
\begin{align*}
\fii&=\phantom{(w\to x)\to(w\to y)\to{}}\bigl((x\to y)\to z\bigr)\to\bigl((y\to x)\to z\bigr)\to z,\\
\psi&=(w\to x)\to(w\to y)\to\bigl((x\to y)\to z\bigr)\to\bigl((y\to x)\to z\bigr)\to z.
\end{align*}
Clearly, $\IPC_{\to,\land}+\psi$ and $\IPC_{\to,\bot}+\psi$ derive~$\fii$ by substituting $x\land y$ or~$\bot$ for~$w$.
However, $\IPC_{\to,\lor}+\psi\nvdash\fii$, as $\psi$ is valid in the algebra~$A$ in the proof of~\ref{item:11} above,
but $\fii$ is not.

Note that $\IPC+\fii=\IPC+(x\to y)\lor(y\to x)$ is just the G\"odel--Dummett logic $\lgc{LC}$, and
one can show $\IPC_{\to,\dots}+\fii=\lgc{LC}_{\to,\dots}$; cf.\ Example~\ref{exm:lc-impl}.
\end{Exm}

\section{Complexity of formulas in intuitionistic logic}\label{sec:cxt-flas}

In this section, we present several lower bounds on formula complexity in intuitionistic fragments. Our main aim is to
show that constraints employed in Section~\ref{sec:elim-land} are necessary. That is, Lemma~\ref{lem:conj-free} tells us
that every formula~$\fii$ can be in~$\IPC$ transformed into a conjunction
\begin{equation}\label{eq:46}
\ET_{i<m}\fii_i
\end{equation}
of $\land$-free formulas
while not introducing any new connectives; for the $\{\to,\land\}$ and $\{\to,\land,\bot\}$ fragments,
this can be done in such a way that $m$ is at most the number of variables (plus $1$ for~$\bot$), and the circuit size
of~$\fii_i$ is polynomial in the circuit size of~$\fii$ (Lemma~\ref{lem:imp-conj-elim}). We will show that $m$ needs to be exponentially large for
formulas with~$\lor$ (Proposition~\ref{thm:moncon}), and that the formula size of~$\fii_i$ may need to be exponentially large
even for $\{\to,\land\}$-formulas (Proposition~\ref{prop:imp-conj} and Corollary~\ref{cor:imp-conj}). In the course of establishing the latter, we
will also prove a linear lower bound on implication nesting depth (Proposition~\ref{prop:int-limd}).

We first deal with disjunctions.
\begin{Prop}\label{thm:moncon}
Let $L\sset\lgc{KC}$ be a si logic, and $\fii$ be a monotone formula. The optimal (in terms of~$m$) expression
of~$\fii$ as a conjunction
\begin{equation}\label{eq:39}
\vdash_L\fii\eq\ET_{i<m}\fii_i
\end{equation}
with $\fii_i$ $\land$-free is the (unique) nonredundant monotone conjunctive normal form of~$\fii$. In particular, the
formula
\[\fii=\LOR_{i<n}(p_i\land q_i)\]
of size~$O(n)$ cannot be written as a conjunction of less than $2^n$ $\land$-free formulas.
\end{Prop}
\begin{Pf}
Assume $\fii$ uses the variables $\{p_i:i<n\}$. Consider the model $\cM=\p{\pw{[n]},{\sset},{\model}}$, where
$[n]=\{0,\dots,n-1\}$, and $I\model p_i$ iff $i\in I$. Since any formula defines an up-set, and each cone $I\up$
in~$\cM$ is definable by $\ET_{i\in I}p_i$, every formula is in~$\cM$ equivalent to a monotone formula.

In fact, we claim that any $\{\to,\lor,\bot\}$-formula $\psi$ is in~$\cM$ equivalent to~$\top$ or to a (possibly empty)
disjunction of variables. This follows by induction on the complexity of~$\psi$: it is enough to check that
\[\cM\model\Bigl(\LOR_{i\in I}p_i\to\LOR_{j\in J}p_j\Bigr)\eq\begin{cases}
  \top&\text{if $I\sset J$,}\\
  \displaystyle\LOR_{j\in J}p_j&\text{otherwise.}
\end{cases}\]
To see this, assume $i\in I\bez J$, and $K\nmodel\LOR_{j\in J}p_j$, which means $J\cap K=\nul$. Then $K'=K\cup\{i\}$
satisfies $K\le K'\model\LOR_Ip_i$, and $K'\nmodel\LOR_Jp_j$, as $J\cap K'=\nul$. Thus,
\[K\nmodel\LOR_{i\in I}p_i\to\LOR_{j\in J}p_j.\]

Now, if \eqref{eq:39} holds, then $\fii$ is in~$\cM$ equivalent to a monotone formula
\[\fii^*=\ET_{i<m}\fii_i^*,\]
where each $\fii_i^*$ is a disjunction of variables (or $\top$, which can be omitted). However, if $\alpha,\beta$ are
monotone formulas, then $\cM\model\alpha\to\beta$ only when $\alpha\to\beta$ is provable (in~$\IPC$, $L$, or $\CPC$, this
is all equivalent): if $v$ is a variable assignment such that $v(\alpha)=1$ and $v(\beta)=0$, then $I=\{i:v(p_i)=1\}$ satisfies
$I\model\alpha$ and $I\nmodel\beta$. Thus,
\[\vdash\fii\eq\fii^*,\]
where $\fii^*$ is a monotone CNF with $m$~clauses.
\end{Pf}
\begin{Rem}
In $\CPC$, any formula is equivalent to a $\{\to,\bot\}$-formula, and any positive formula~$\fii$ can be written as a
conjunction of $m$~$\{\to\}$-formulas, where $m$ is the minimal length of a conjunction of variables that implies~$\fii$
(for monotone~$\fii$, this is the length of the shortest conjunction in its monotone DNF): indeed, if $\ET_{i\in I}p_i$
implies~$\fii$, then
\[\fii\eq\ET_{i\in I}(p_i\lor\fii),\]
and a Boolean function can be expressed by a $\{\to\}$-formula iff it majorizes a
variable.
\end{Rem}

Now we turn to $\{\to,\land\}$-formulas. While every $\{\to,\land\}$-circuit is equivalent to a polynomially larger
conjunction of $\{\to\}$-circuits, we will prove that a $\{\to,\land\}$-formula may need exponential size when
expressed as a conjunction of $\land$-free formulas~\eqref{eq:46}.

First however, it is not hard to see that if $\fii$ is a $\{\to,\land\}$-formula of depth~$d$, we can make
the size of~\eqref{eq:46} to be $\lh\fii^{O(d)}$; this is a quasipolynomial bound in the ideal case when $d=O(\log\Abs\fii)$. Thus, if we
want exponential lower bounds on the size of~\eqref{eq:46}, we should better make sure that $\fii$ cannot be written by a
formula of low depth.

In $\CPC$, every formula can be transformed into a polynomially larger formula of logarithmic
depth by Spira's lemma. In contrast, we will show that in~$\IPC$, formulas may require linear depth, even using a
rather relaxed measure---we only count the nesting of implications on the left:
\begin{gather*}
\limd(p)=\limd(\bot)=\limd(\top)=0,\\
\limd(\fii\land\psi)=\limd(\fii\lor\psi)=\max\{\limd(\fii),\limd(\psi)\},\\
\limd(\fii\to\psi)=\max\{\limd(\fii)+1,\limd(\psi)\}.
\end{gather*}
Let $U_n=\{\fii:\limd(\fii)\le n\}$; $U_1$-formulas are also called NNIL. These formula classes were studied
in~\cite{nnil}. There are only finitely many $U_n$-formulas up to equivalence over a finite set of variables.

If $\cM=\p{M,{\le},{\model}}$ is a Kripke model for a finite set of variables~$V$, and $x,y\in M$, let us write
\[x\to_ny\iff\forall\fii\in U_n\,(x\model\fii\TO y\model\fii).\]
Notice that $\to_n$ is transitive, and contains~$\le$. Clearly,
\[x\to_0y\iff\forall p\in V\,(x\model p\TO y\model p).\]
By a variant of a characterization from~\cite{nnil}, we have $x\to_ny$ if and only if there is a family of
relations $\p{{\TO}_k:k\le n}$ such that
\begin{enumerate}
\item\label{item:19} $x'\TO_ny$ for some $x'\ge x$;
\item $u\TO_kv$ implies $u\to_0v$;
\item $u\TO_{k+1}v$ implies $v\TO_ku$;
\item\label{item:20} whenever $u\TO_{k+1}v\le v'$, there is $u'$ such that $u\le u'\TO_{k+1}v'$.
\end{enumerate}
(The reader should think of~$u\TO_kv$ as an under-approximation of $u\to_kv\to_{k-1}u$.)

Let us define the formula
\begin{equation}\label{eq:47}
\alpha_n(p_0,\dots,p_n)=(\cdots((p_0\to p_1)\to p_2)\to\cdots)\to p_n
\end{equation}
for $n\ge0$. That is, $\alpha_0=p_0$, and $\alpha_{n+1}=(\alpha_n\to p_{n+1})$.
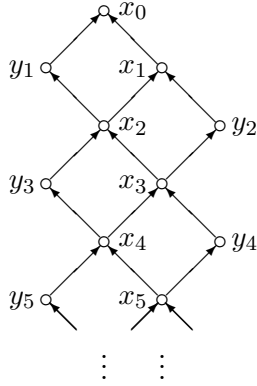
\begin{figure}
\centering
\unitlength=1em
\begin{picture}(10,14)

\cput(4,13){$\R$}
\put(4.5,13.1){$x_0$}

\put(2.15,11.4){\vector(1,1){1.73}}
\put(5.85,11.4){\vector(-1,1){1.73}}

\cput(2,11){$\R$}
\put(.7,11.1){$y_1$}
\cput(6,11){$\R$}
\put(4.5,11.1){$x_1$}

\put(3.85,9.4){\vector(-1,1){1.73}}
\put(4.15,9.4){\vector(1,1){1.73}}
\put(7.85,9.4){\vector(-1,1){1.73}}

\cput(4,9){$\R$}
\put(4.5,9.1){$x_2$}
\cput(8,9){$\R$}
\put(8.4,9.1){$y_2$}

\put(2.15,7.4){\vector(1,1){1.73}}
\put(5.85,7.4){\vector(-1,1){1.73}}
\put(6.15,7.4){\vector(1,1){1.73}}

\cput(2,7){$\R$}
\put(.7,7.1){$y_3$}
\cput(6,7){$\R$}
\put(4.5,7.1){$x_3$}

\put(3.85,5.4){\vector(-1,1){1.73}}
\put(4.15,5.4){\vector(1,1){1.73}}
\put(7.85,5.4){\vector(-1,1){1.73}}

\cput(4,5){$\R$}
\put(4.5,5.1){$x_4$}
\cput(8,5){$\R$}
\put(8.4,5.1){$y_4$}

\put(2.15,3.4){\vector(1,1){1.73}}
\put(5.85,3.4){\vector(-1,1){1.73}}
\put(6.15,3.4){\vector(1,1){1.73}}

\cput(2,3){$\R$}
\put(.7,3.1){$y_5$}
\cput(6,3){$\R$}
\put(4.5,3.1){$x_5$}

\put(3.05,2.2){\vector(-1,1){.93}}
\put(4.95,2.2){\vector(1,1){.93}}
\put(7.05,2.2){\vector(-1,1){.93}}

\cput(4,.5){$\vdots$}
\cput(6,.5){$\vdots$}
\end{picture}
\caption{Kripke frame used in Proposition~\ref{prop:int-limd}}
\label{fig:frame}
\end{figure}
\begin{Prop}\label{prop:int-limd}
If $L$ is a si logic included in $\lgc{KC+BW_2}$, and $n\ge1$, the implicational $U_n$-formula~$\alpha_n$
is not equivalent over~$L$ to any $U_{n-1}$-formula.
\end{Prop}
\begin{Pf}
We consider the model $\cM=\p{M,{\le},{\model}}$, where
\begin{gather*}
M=\{x_n:n\ge0\}\cup\{y_n:n\ge1\},\\
x_n<x_{n-1},\quad x_{n+1}<y_n<x_{n-1},\\
y_1\model p_0,\\
x_n,y_n\model p_{n+1}.
\end{gather*}
(See Figure~\ref{fig:frame}.) A straightforward induction on~$n$ shows
\[u\model\alpha_n\iff u\ge y_{n+1},\]
thus the result will follow if we prove $y_{n+1}\to_{n-1}x_{n+1}$ (in fact, also $x_{n+1}\to_ny_{n+1}$). To this end, it
is enough to define relations $\TO_n$ for $n\in\omega$ by
\[{\TO_n}={\id}_M\cup\{\p{x_m,y_m}:m\ge n+1\}\cup\{\p{y_m,x_m}:m\ge n+2\},\]
and observe that the conditions \ref{item:19}--\ref{item:20} from the characterization of~$\to_n$ are satisfied.
\end{Pf}

\begin{Def}
With every $\{\to,\bot\}$-formula~$\fii(\vec p)$, we associate a rooted tree~$T_\fii$ whose nodes are labelled by
variables~$p_i$ or by~$\bot$ (these will be collectively called \emph{atoms}) using the following inductive definition: if
\[\fii=\fii_0\to\dots\to\fii_{n-1}\to v,\]
where $n\ge0$, and $v$ is an atom (i.e., $v=H(\fii)$), then $T_\fii$ has a root labelled by~$v$ with children $T_{\fii_0}$, \dots,
$T_{\fii_{n-1}}$.

More generally, we associate each $\land$-free formula with a labelled forest~$T_\fii$: if $\fii$ is an atom, then
$T_\fii$ is the singleton tree labelled~$\fii$; if $\fii=\psi\lor\chi$, then $T_\fii$ is the
disjoint union of $T_\psi$ and~$T_\chi$; if $\fii=\psi\to\chi$, then $T_\fii$ is obtained from~$T_\chi$ by attaching
copies of all trees from~$T_\psi$ as children to roots of all trees of~$T_\chi$. For example, this is $T_{p\lor q\to
r\lor s}$:
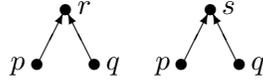
\begin{center}
\unitlength=1em
\begin{picture}(9,3)

\cput(2,2.2){$\I$}
\put(2.4,2.3){$r$}

\put(1.1,0.7){\vector(1,2){.82}}
\put(2.9,0.7){\vector(-1,2){.82}}

\cput(1,0.3){$\I$}
\put(0.1,0.4){$p$}
\cput(3,0.3){$\I$}
\put(3.4,0.4){$q$}

\cput(7,2.2){$\I$}
\put(7.4,2.3){$s$}

\put(6.1,0.7){\vector(1,2){.82}}
\put(7.9,0.7){\vector(-1,2){.82}}

\cput(6,0.3){$\I$}
\put(5.1,0.4){$p$}
\cput(8,0.3){$\I$}
\put(8.4,0.4){$q$}
\end{picture}
\end{center}

A \emph{path} in~$T_\fii$ is a sequence of labels of nodes $x_0,\dots,x_n\in T_\fii$, where $x_{i+1}$
is the parent of~$x_i$ for each~$i<n$, and $x_n$ is a root.
\end{Def}
\begin{Lem}\label{lem:paths}
Let $\cM$ be the Kripke model considered in the proof of Proposition~\ref{prop:int-limd}. Put $\beta_n=\alpha_n\to p_n$ and
$\gamma_n=\alpha_n\land p_{n+1}$ for $n\ge0$; notice that in~$\cM$, $\beta_0\equiv\top$, $\beta_1\equiv\alpha_0\equiv p_0$, and
$\gamma_0\equiv p_1$.

Every $\{\to,\lor\}$-formula is in~$\cM$ equivalent to an implicational formula, namely $\top$, $p_n$, $\alpha_n$,
or~$\beta_n$ for some~$n\ge1$.
A $\land$-free formula is equivalent to an implicational formula as above, or to~$\bot$.
An arbitrary formula is equivalent either to one of the above mentioned, or to $\gamma_n$ for some $n\ge1$.

If $\fii$ is a $\{\to,\bot\}$-formula not equivalent to~$\top$, then $T_\fii$ has a path of the following form,
depending on the canonical equivalent of~$\fii$:
\begin{center}
\renewcommand\arraystretch{1.2}
\begin{tabular}{ll}
$p_n$, $n\ge1$&$p_n$\\
$\bot$&$\bot$\\
$p_0\equiv\alpha_0\equiv\beta_1$&$p_0p_1^{2k}$\\
$\alpha_n$, $n\ge1$&$p_0p_1^{2k_1+1}\dots p_n^{2k_n+1}$\\
$\beta_n$, $n\ge2$&$p_0p_1^{2k_1+1}\dots p_{n-1}^{2k_{n-1}+1}p_n^{2k_n+2}$
\end{tabular}
\end{center}
The same also holds for $\land$-free formulas $\fii$ not equivalent to $\top$ or $\beta_n$ for~$n\ge2$.
\end{Lem}
\begin{Pf}
Recall that $u\model\alpha_n$ iff $u\ge y_{n+1}$. Also, we have $u\model\beta_n$ iff $u\ge y_n$ or (for $n>1$) $u\ge
y_{n-1}$; and $u\model\gamma_n$ iff $u\ge x_n$.

Let $F=\{\top,p_n,\alpha_n,\beta_n:n\ge1\}$, and $F^\bot=F\cup\{\bot\}$. In~$\cM$, every negated formula is equivalent
to $\top$ or~$\bot$, and for all $n\ge1$, $\cM$ satisfies
\begin{equation}\label{eq:40}
p_n\to\beta_n,\quad\beta_n\to p_{n+1},\quad p_n\to\alpha_n,\quad\alpha_n\to\beta_{n+1}.
\end{equation}
Using this, it is not hard to verify that for any $\psi,\chi\in F^\bot$, $\psi\to\chi$ is equivalent to $\top$
or~$\chi$, and $\psi\lor\chi$ is equivalent to $\psi$ or~$\chi$, with only the following exceptions,
where $n\ge1$ ($n>1$ in the cases involving $n-1$):
\begin{equation}\label{eq:48}
\left.
\begin{gathered}
(\alpha_{n-1}\to p_n)\equiv(\beta_n\to p_n)\equiv\alpha_n,\\
(\alpha_n\to p_n)\equiv\beta_n,\\
(\alpha_n\lor p_{n+1})\equiv(\alpha_n\lor\beta_n)\equiv(\alpha_n\lor\alpha_{n-1})\equiv\beta_{n+1}.
\end{gathered}
\qquad\right\}
\end{equation}
This shows that $F$ and~$F^\bot$ are closed under $\to$
and~$\lor$, hence every $\{\to,\lor\}$-formula is equivalent to a formula from~$F$, and every
$\{\to,\lor,\bot\}$-formula to a formula from~$F^\bot$. A general formula is equivalent to a conjunction of
$\land$-free formulas; using~\eqref{eq:40}, the only cases where the conjunction of elements of $F^\bot$ is different
from one of the conjuncts are $\alpha_n\land\beta_n\equiv p_n$, and $\alpha_n\land\alpha_{n+1}\equiv\alpha_n\land
p_{n+1}=\gamma_n$.

Assume $\fii$ is an $\{\to,\bot\}$-formula not equivalent to~$\top$. We can simplify~$\fii$ by repeatedly replacing
subformulas $\psi\to\chi$ with~$\chi$ when equivalent, yielding a new formula~$\fii'$; this transformation amounts to removing the
subtree $T_\psi$ from~$T_\fii$, hence any path in~$T_{\fii'}$ is also a path in~$T_\fii$. Notice that in
particular, all subformulas $\psi$ equivalent to~$\top$ get deleted. By~\eqref{eq:48},
$\fii'$ contains no nontrivial formulas equivalent to $p_n$ for~$n\ge1$ or to~$\bot$, and the only implications left
in~$\fii'$ have variables on the 
right-hand side, i.e., $T_{\fii'}$ consists of a single path. It is then easy to check that this path must have the form
given in the table.

Now, let $\fii$ be an $\{\to,\lor,\bot\}$-formula. We simplify it by replacing subformulas $\psi\to\chi$ with~$\chi$
when equivalent (as above), and by replacing $\psi\lor\chi$ with $\psi$ or~$\chi$ when this does not change $\fii$ (up
to equivalence). Let $\fii'$ denote the resulting formula. This transformation again amounts to removal of
several subtrees from the forest~$T_\fii$, hence every path in~$T_{\fii'}$ is a path in~$T_\fii$.
As before, $\fii'$ contains no nontrivial subformulas equivalent to $p_n$ for~$n\ge1$, or
to~$\bot$; in particular, if $\fii'$ itself is equivalent to one of those, it consists of a single symbol.

Assume $\fii'$ is equivalent to~$\alpha_n$; we will show that $T_{\fii'}$ is a path as given in the table by induction
on~$\lh{\fii'}$. By~\eqref{eq:48}, there is no nontrivial way to express $\alpha_1$
or~$\alpha_0=\beta_1$ as a disjunction; this easily gives the result for $n=0,1$ as in the $\{\to,\bot\}$ case. Assume
$n\ge2$. By~\eqref{eq:48}, $\fii'$ has one of the forms
\begin{gather*}
\psi\to p_n,\\
(\fii''\to p_n)\to p_n,\\
\psi\lor\chi\to p_n,
\end{gather*}
where $\psi\equiv\alpha_{n-1}$, $\fii''\equiv\alpha_n$, $\chi$ is equivalent to $p_n$, $\beta_{n-1}$, or~$\alpha_{n-2}$,
and the disjunction may be written in reverse order. In the first two
cases, it suffices to apply the induction hypothesis to $\psi$ or~$\fii''$ (resp.). The third case is actually
impossible, as we could simplify the formula by deleting the ``${}\lor\chi$'' part.
\end{Pf}

\begin{Def}\label{def:unnested}
Let $\fii$ be a $\land$-free formula. An occurrence of a subformula $\psi$ in~$\fii$ is \emph{unnested} if it is not
within the left scope of an implication. That is, the set $U(\fii)$ of unnested occurrences of subformulas of~$\fii$ can be
defined inductively by
\begin{align*}
U(v)&=\{v\}\qquad\text{if $v$ is an atom,}\\
U(\fii\to\psi)&=\{\fii\to\psi\}\cup U(\psi),\\
U(\fii\lor\psi)&=\{\fii\lor\psi\}\cup U(\fii)\cup U(\psi).
\end{align*}
Notice that atoms in~$U(\fii)$ are exactly the labels of roots of the trees in~$T_\fii$.
\end{Def}
\pagebreak[2]
\begin{Lem}\label{lem:unnested}
Let $\fii$ be a $\land$-free formula.
\begin{enumerate}
\item\label{item:22}
If $\top\in U(\fii)$, then $\fii\equiv\top$.
\item\label{item:21}
If $T_\fii$ has a path~$Pv$, where $v$ is an atom, then $v\in U(\fii)$. If moreover $P$ is nonempty, then $\fii$ has an
unnested subformula of the form $\psi\to\chi$ such that $v\in U(\chi)$, and $T_\psi$ has a path~$P$.
\end{enumerate}
\end{Lem}
\begin{Pf}
By induction on the complexity of~$\fii$.
\end{Pf}

For simplicity, let us define the size of a formula as the number of symbols in Polish notation, that is, the number of
occurrences of all variables and connectives, disregarding auxiliary characters (brackets, variable indices).
\begin{Prop}\label{prop:imp-conj}
If $L$ is a si logic included in~$\lgc{KC+BW_2}$, then any $\land$-free formula $L$-equivalent to
\[\alpha^{\land2}_n=\alpha_n(p_{0,0}\land p_{0,1},p_{1,0}\land p_{1,1},\dots,p_{n-1,0}\land p_{n-1,1},p_n)\]
must have size at least~$2^{n+2}-3$.
\end{Prop}
\begin{Pf}
Let $\fii$ be a $\land$-free formula equivalent to $\alpha_n^{\land2}$. For any $f\colon[n]\to\{0,1\}$, we consider the
substitution
\begin{align*}
\sigma_f(p_{i,f(i)})&=p_i,\\
\sigma_f(p_{i,1-f(i)})&=\top.
\end{align*}
Notice that $\sigma_f(\fii)\equiv\sigma_f(\alpha_n^{\land2})\equiv\alpha_n$. Let $\fii_f$ denote the formula
$\sigma_f(\fii)$ with subformulas made true by~$\sigma_f$ deleted. By Lemma~\ref{lem:paths}, $T_{\fii_f}$ has a path of
the form $p_0p_1^{2k_1+1}\dots p_n^{2k_n+1}$ for some $k_1,\dots,k_n\ge0$ (depending on~$f$). Lifting it back to~$T_\fii$, and
using Lemma~\ref{lem:unnested}, we can find a sequence of occurrences of subformulas
\begin{equation}\label{eq:49}
\fii=\psi_{f,n,0}\Sset\fii_{f,n,0}\Sset\psi_{f,n,1}\Sset\dots\Sset\fii_{f,n,2k_n}\Sset\psi_{f,n-1,0}\Sset\dots\Sset\fii_{f,1,2k_1}\Sset\psi_{f,0,0}\Sset\fii_{f,0,0}
\end{equation}
such that:
\begin{enumerate}
\item $\fii_{f,i,j}\in U(\psi_{f,i,j})$.
\item $\fii_{f,i+1,j}$ is an implication with premise $\psi_{f,i+1,j+1}$ for $j<2k_{i+1}$, or $\psi_{f,i,0}$ for $j=2k_{i+1}$.
\item $p_{i,f(i)}$ has an unnested occurrence (denoted~$v_{f,i,j}$) in~$\fii_{f,i,j}$. Here, ``$p_{n,f(n)}$'' stands
for~$p_n$.
\item\label{item:25}
There are no variables in $U(\psi_{f,i,j})$ other than $p_{0,f(0)},\dots,p_{n,f(n)}$.
\end{enumerate}
\begin{Cl}
Let $f,g\colon[n]\to\{0,1\}$, and $i\le n$.
\begin{enumerate}
\item\label{item:23}
If $i<n$ and $f(i)\ne g(i)$, then $\fii_{f,i,0}\nsset\fii_{g,i,0}$.
\item\label{item:24}
$v_{f,i,0}\ne v_{g,i',0}$ unless $i=i'$, and $f$ agrees with~$g$ on $\{i,\dots,n-1\}$.
\end{enumerate}
\end{Cl}
\begin{Pf*}

\ref{item:23}: Assume that $\fii_{f,i,0}$ is a suboccurrence of~$\fii_{g,i,0}$. Since each $\psi_{f,i',j'}$ (except
$\psi_{f,n,0}$) has an appropriate $\fii_{f,i'',j''}$ formula as its parent, it follows
from~\eqref{eq:49} that $\fii_{f,i',j'}\sset\fii_{g,i,0}\sset\psi_{f,i',j'}$ for some $i'\ge i$ and $j'\le2k_{i'}$.
However, since $\fii_{f,i',j'}$ is unnested in~$\psi_{f,i',j'}$, so must be~$\fii_{g,i,0}$, hence 
$p_{i,g(i)}$ has an unnested occurrence in~$\psi_{f,i',j'}$. This contradicts \ref{item:25} above.

\ref{item:24}: Assume $v_{f,i,0}=v_{g,i',0}$. Since $v_{f,i,0}$ is an occurrence of~$p_{i,f(i)}$, we must have $i=i'$, and
$f(i)=g(i)$. If $f(i'')$ were different from~$g(i'')$ for some $i''>i$, then $\fii_{f,i'',0}$ and~$\fii_{g,i'',0}$
would be incomparable by the first part of the claim, hence so would be their respective subformulas $v_{f,i,0}$
and~$v_{g,i,0}$.
\end{Pf*}

The claim implies that $\bigl\{v_{f,i,0}:f\colon[n]\to\{0,1\},i\le n\bigr\}$ consists of at least
$2^n+\dots+2^0=2^{n+1}-1$ distinct occurrences of variables in~$\fii$, hence the total size of~$\fii$ is at least
$2^{n+2}-3$.
\end{Pf}
\begin{Rem}
The bound is tight: we can express $\alpha_n^{\land2}$ by an implicational formula of size $2^{n+2}-3$ using the recurrence
\begin{align*}
\alpha_0^{\land2}&=p_0,\\
\alpha_{n+1}^{\land2}&=\alpha_n^{\land2}(p_n/p_{n,0})\to\alpha_n^{\land2}(p_n/p_{n,1})\to p_{n+1}.
\end{align*}
\end{Rem}
\begin{Cor}\label{cor:imp-conj}
If $L$ is a si logic included in~$\lgc{KC+BW_2}$, and $\fii$ is a conjunction of $\land$-free formulas $L$-equivalent to
\[\xi_2(\alpha_n)=\alpha_n(p_{0,0}\land p_{0,1},p_{1,0}\land p_{1,1},\dots,p_{n,0}\land p_{n,1}),\]
then $\lh\fii\ge2^{n+3}-5$.
\end{Cor}
\begin{Pf}
If $\fii=\ET_{i<m}\fii_i$, then $\alpha_{n+1}^{\land2}$ is equivalent to the formula $\fii_0\to\dots\to\fii_{m-1}\to
p_{n+1}$ of size $\lh\fii+2$.
\end{Pf}
\begin{Exm}\label{exm:rn}
We may interpret Proposition~\ref{prop:imp-conj} as an explicit exponential separation between formula and circuit size in a fragment of
intuitionistic logic, and the reader may wonder whether there is such a separation also in the full intuitionistic
language. This is indeed true, and essentially well known. One can take for example the Rieger--Nishimura formulas
\begin{gather*}
\M{rn}_0=\bot,\qquad
\M{rn}_1=p,\qquad
\M{rn}_2=\neg p,\\
\M{rn}_{2n+3}=\M{rn}_{2n+1}\lor\M{rn}_{2n+2},\\
\M{rn}_{2n+4}=\M{rn}_{2n+2}\to\M{rn}_{2n+1}
\end{gather*}
(see \cite[Exm.~7.66]{cha-zax}). Every intuitionistic formula in one variable $\fii(p)$ is equivalent to $\top$,
or to $\M{rn}_k$ for some~$k$, and it is not difficult to estimate $k=O(\log\Abs\fii)$. On the other hand, $\M{rn}_k$
is defined by a circuit of size~$O(k)$. Thus, \emph{every intuitionistic formula in one variable can be expressed by an
exponentially smaller circuit.} (In a sense, something even stronger holds in classical logic, as every formula in one
variable has circuit size~$O(1)$. However, the intuitionistic bound is meaningful as unlike~$\CPC$, there are
infinitely many nonequivalent formulas in one variable in~$\IPC$.)
\end{Exm}
\end{document}